\documentclass[lettersize,journal]{IEEEtran}
\usepackage{amsmath,amsfonts}
\usepackage{array}
\usepackage{textcomp}
\usepackage{stfloats}
\usepackage{url}
\usepackage{verbatim}
\usepackage{graphicx}
\usepackage{cite}
\usepackage{xcolor}
\usepackage{mathrsfs}
\usepackage{amssymb}
\usepackage{stfloats}
\usepackage{amsmath}
\usepackage[linesnumbered,ruled,vlined]{algorithm2e}
\usepackage{subcaption}
\usepackage{subcaption}

\DeclareCaptionLabelFormat{singleparen}{#2} 
\makeatletter
\renewcommand\p@subfigure{\thefigure}
\makeatother

\begin{document}
\def\IEEEtitletopspace{0pt} 
\IEEEaftertitletext{\vspace{-15pt}}

\title{Multi-Static ISAC Assisted by Double-Side Fluid Antenna System}

\author{Qinyuan Zheng, Pengcheng Zhu,~\IEEEmembership{Member,~IEEE},
\thanks{Qinyuan Zheng, and Pengcheng Zhu are with the National Mobile Communications Research Laboratory, Southeast University, Nanjing 210096, China (e-mail:zhengqy@seu.edu.cn; p.zhu@seu.edu.cn).}}

\markboth{Journal of \LaTeX\ Class Files,~Vol.~14, No.~8, August~2021}%
{Shell \MakeLowercase{\textit{et al.}}: A Sample Article Using IEEEtran.cls for IEEE Journals}


\maketitle

\begin{abstract}
As a pivotal usage scenario for 6G networks, integrated sensing and communication (ISAC) has emerged as a focal point of both academic and industrial research. To accommodate the heterogeneous connectivity requirements of future networks while jointly enhancing both the sensing and communication performance, this paper integrates the multi-static ISAC architecture with double-side fluid antenna system (DS-FAS) to fully exploit the available spatial degrees-of-freedom. Specifically, we establish a joint optimization framework for FA positions and transmit beamforming to maximize the target detection probability while satisfying the communication quality-of-service requirements. Recognizing the intricate coupling between the double-side FA positions and transmit beamforming, instead of trying to obtain an initial feasible point, we resort to the penalty-based mechanism to ensure the robustness against initial feasibility without introducing additional non-convexity. An alternating optimization-based algorithm is proposed to solve the decoupled subproblems. Specifically, the transmit beamforming is globally optimized via the semidefinite relaxation technique, while the transmit FA positions are determined using the majorization-minimization method. Finally, leveraging the analyzed FA mechanism, the feasibility subproblem for receive FA positions is transformed into a signal-to-interference-plus-noise ratio maximization one, solved efficiently via a gradient ascent-based approach, which yields superior performance over the feasibility-based benchmark with reduced complexity. Numerical results demonstrate the superiority of the considered DS-FAS-assisted multi-static ISAC systems in both noise-limited and interference-limited scenarios, while key insights for practical deployment are further extracted from the simulation analysis.

\end{abstract}

\begin{IEEEkeywords}
Integrated sensing and communication, multi-static ISAC, fluid antenna system, target detection.
\end{IEEEkeywords}

\vspace{0pt}
\section{Introduction}

\subsection{Background}
    \IEEEPARstart{T}{he} advancement of wireless sensing and communication (S\&C) technologies has led to the emergence of diverse applications with more stringent performance requirements, placing immense pressure on fundamental resources such as time, frequency, power, and space~\cite{you2021towards}. To enhance resource efficiency, the once-independent S\&C functionalities are proposed to be realized in the same infrastructure~\cite{10664619}, which is termed as integrated sensing and communication (ISAC). Thanks to the extra degrees-of-freedom (DoFs) provided by multi-antenna technology, the space-division-based ISAC enables directional beamforming to not only suppress interference between communication users but also distinguish between different targets in the same time-frequency resource~\cite{9737357}, which constitutes the fundamental framework of current ISAC. To date, further exploiting the spatial DoFs to meet even higher performance requirements has become essential for the evolution of future ISAC.

    From the system architecture perspective, mono-static ISAC, where the transmitter and receiver are co-located, has been widely investigated~\cite{9124713,10227884,10143420}. As a step further, bi-static ISAC has been proposed to separate the transmitter and receiver~\cite{10147248,10138084}, thereby circumventing the severe self-interference. However, these architectures fail to fully exploit the potential spatial DoFs due to the limited number of transmission and sensing observation points. Inspired by the concept of cell-free architecture in the realm of wireless communication, multi-static ISAC has emerged as a promising solution, where multiple access points (APs) are distributed across the considered area to serve as ISAC transmitters or sensing receivers. By leveraging joint transmission and reception signal processing, substantial macro-diversity and multi-aspect sensing gains can be achieved~\cite{10726912}.

    From the antenna technology perspective, antenna configuration has evolved from conventional fixed-position antennas (FPA) to antennas that can flexibly change their positions within a given area to further unleash the potential spatial DoFs~\cite{11247926}. The concept of "fluid antenna system (FAS)" was first introduced to the field of wireless communication by Wong \textit{et al.}  to represent "position-flexible and shape-flexible antennas" in~\cite{wong2020fluid}, where software-controlled, movable conductive fluid was utilized as antenna element at the user equipment (UE) to attain communication capacity enhancement. Shortly thereafter, another concept termed "movable antenna", whose movement is typically realized via stepper motors, was investigated for wireless communications by Zhu \textit{et al.} in ~\cite{10318061,10243545}. Despite their different origins and physical implementations, these two concepts are unified under the umbrella of position-flexible antennas~\cite{zhu2024historical}. For the sake of clarity and without loss of generality, we shall adopt the term "FAS" throughout the remainder of this paper.

\vspace{0pt}
\subsection{Related Work}
To fully leverage the available spatial DoFs for enhancing both S\&C performance, not only multi-static ISAC~\cite{10032141,10494224,10639146,10516289,10605793,10731920,10746496} but also FAS-assisted ISAC~\cite{10705114,10839251,10696953,10962171,11033708,zhang2025fluidantennaaidedrobustsecure, 10707252,11086422,11353414,11249714,11142587,11075964} have been extensively investigated. The related works will be summarized in this subsection to finally motivate our work.

\textbf{1) Studies on Multi-Static ISAC:}
The efficacy of multi-static ISAC has been validated across a variety of sensing scenarios~\cite{10032141,10494224}. Specifically, target localization was investigated in~\cite{10032141}, where a Pareto optimization framework was proposed to characterize the performance tradeoff between S\&C. Furthermore, the authors in~\cite{10494224} developed a posterior ratio test detector for target detection, and a power allocation algorithm was devised to maximize the sensing signal-to-interference-plus-noise ratio (SINR) while satisfying the quality-of-service (QoS) requirements of UEs. Moving towards more practical implementations, various issues such as scalability of multi-static architectures~\cite{10639146}, imperfect channel state information~\cite{10516289}, and physical layer security (PLS)~\cite{10605793} have been addressed in the context of multi-static ISAC systems. Building on these foundations, emerging technologies such as non-orthogonal multiple access~\cite{10731920}, rate-splitting multiple access (RSMA)~\cite{10032141}, and reconfigurable intelligent surface (RIS)~\cite{10746496} have been integrated into multi-static ISAC systems. These integrations have been demonstrated to provide greater flexibility in exploiting available spatial DoFs, thereby enhancing both S\&C performance.

\textbf{2) Studies on ISAC Assisted by Single-Side FAS:}
Beyond multi-static ISAC architectures, FAS has been preliminarily integrated at either transmitter or receiver side to further exploit spatial DoFs for ISAC~\cite{10705114,10839251,10696953,10962171,11033708,zhang2025fluidantennaaidedrobustsecure}. At the ISAC transmitter, antenna positions have been selected from predefined discrete ports distributed along a linear space in~\cite{10705114} to achieve energy-efficient ISAC operation. Taking this a step further, the work in~\cite{10839251} considered an ISAC transmitter with antennas moving within a continuous region and an efficient gradient ascent (GA)-based method was proposed to optimize the antenna positions. Regarding the receiver side, the work in~\cite{10696953} considered both UEs and sensing receiver equipped with FAs, aiming to minimize the Cram\'{e}r-Rao bound (CRB) while satisfying the QoS requirements of the UEs. Integrated with RIS, FAs equipped at the ISAC transmitter and receiver sides have been investigated separately in~\cite{10962171} and~\cite{11033708}, respectively, revealing a superimposed effect of these two technologies in exploiting spatial DoFs. To acquire a similar superimposed effect, the work in~\cite{zhang2025fluidantennaaidedrobustsecure} considers UEs equipped with FAs in conjunction with RSMA to enhance the PLS.

\textbf{3) Studies on ISAC Assisted by Double-Side FAS:}
Motivated by the promising performance of Single-Side FAS (SS-FAS)-assisted ISAC, it is natural to envisage an ISAC system where both the transmitter and receiver sides are equipped with FAs to unlock ultimate potential of spatial DoFs~\cite{10707252,11086422,11353414,11249714,11142587,11075964}. Focusing on a simplified scenario involving one UE and one target, the work in~\cite{10707252} co-designed the transmit beamforming and positions of both transmit and receive antennas to obtain a local optimal solution. Within a similar system setup, global optimal solution and efficient algorithm for local optimal solution have been proposed in~\cite{11086422} for cases with either transmit-only or receive-only FAs, and the double-side FAS (DS-FAS) case can be addressed by integrating the aforementioned methods. Further extending this to a multi-UE ISAC scenario, a reinforcement learning-based method was proposed in~\cite{11353414} to minimize the worst case CRB while satisfying the UEs' QoS requirements. Then, full-duplex scenarios, where both sides are equipped with FAs, have been investigated. Specifically, FAs with discrete positions or continuous regions were separately studied in~\cite{11249714} and~\cite{11142587}, respectively. Beyond traditional far-field ISAC systems, near-field channel conditions were investigated in~\cite{11075964}, where the performance gains offered by DS-FAS were demonstrated.

\vspace{0pt}
\subsection{Research Gap and Main Contributions}
Inspired by the enhanced spatial DoFs inherent in multi-static ISAC architecture and DS-FAS, this paper integrates these two paradigms into a sensing-centric ISAC scenario. Specifically, we consider a framework where multiple ISAC transmitters and sensing receivers collaboratively perform joint target detection while simultaneously delivering information to UEs, which has not been fully investigated in the existing literature. While the benefits of FAS have been widely acknowledged, little attention has been paid to the initialization of FA positions in prior works. Although several simple initialization schemes, such as the circle packing (CP) method~\cite{10243545} or uniform linear array (ULA) configurations~\cite{10696953}, have been adopted, these heuristic approaches fail to guarantee that the stringent communication or sensing performance requirements are satisfied at the initial stage, which may hinder subsequent optimization procedures. A more sophisticated method was proposed in~\cite{10839251}, where the FA positions are determined by traversing all predefined discrete points. However, its computational complexity grows exponentially with both the number of FAs per AP and the total number of APs, rendering it impractical for the multi-static ISAC scenario considered in this paper. Therefore, it becomes imperative to either develop an effective initialization strategy or design an optimization algorithm that is inherently robust to the initial feasibility of the problem. Besides the above issue, the potential of UE-side FA in sensing-centric ISAC systems remains under-exploited. Current studies, such as~\cite{10696953}, merely treat the optimization of UE-side FA positions as a feasibility problem, which may limit the performance gains of the considered DS-FAS-assisted ISAC systems. Motivated by the novel scenario, our objective is to jointly optimize the FA positions at both the transmitters and receivers, alongside the transmit beamforming, to maximize the detection probability while adhering to the UEs' QoS requirements. Furthermore, the aforementioned two issues will also be addressed in the newly developed algorithm. Our main contributions are summarized as follows:
\begin{itemize}
    \item 
    \textbf{DS-FAS-assisted ISAC with GLRT Detector:} A DS-FAS-assisted multi-static ISAC framework is proposed where both the ISAC transmitters and sensing receivers are equipped with multiple FAs, and each UE is equipped with a single FA. All FAs are capable of moving within a 2D region to fully exploit the available spatial DoFs. To characterize the target detection performance, a generalized likelihood ratio test (GLRT) detector is applied and the closed-form detection probability is derived as functions of both FAs positions and transmit beamforming.

    \item 
    \textbf{Initial Feasibility-Robust Problem Formulation:} A target detection probability maximization problem is formulated via jointly optimizing the FA positions and transmit beamforming. Distinct from traditional designs, a penalty-based mechanism is integrated to render the formulation robust against initial feasibility. Furthermore, through tailored transformation, the introduced slack variable will not impose additional non-convexity to the constraints. Building on the formulated problem, the sensing-only scenario is further analyzed to provide more insights into the FA mechanism within the considered system, which heuristically motivates the design of the subsequent optimization algorithm.

    \item 
    \textbf{AO-based Joint Optimization Algorithm:} An alternating optimization (AO) framework is developed to decouple the variables to be optimized into three blocks. Specifically, for transmit beamforming, the semidefinite relaxation (SDR) technique is employed to obtain a globally optimal solution for the corresponding subproblem. Subsequently, the transmit FA positions are optimized via the majorization-minimization (MM) method. Finally, drawing on the previously analyzed FA mechanism, we reformulate the feasibility subproblem for receive FA positions to a signal-to-interference-plus-noise ratio (SINR) maximization problem. This transformation enables the problem to be efficiently solved for each UE in parallel via a gradient ascent (GA)-based approach, which has much lower computational complexity but can achieve higher performance gains.

    \item 
    \textbf{Performance Verification:} Extensive simulation results are presented to validate the superiority of the proposed DS-FAS framework across both noise-limited and interference-limited scenarios. Specifically, significant performance gains in terms of detection probability and achievable communication QoS can be observed compared to various benchmark schemes. Furthermore, several suggestions for practical deployments are extracted from the simulation analysis. Finally, the impact of the moving region sizes of both sides is also demonstrated.
    
\end{itemize}
\vspace{0pt}

\section{System Model}
\begin{figure}[htbp] 
    \centering 
    \includegraphics[width=0.35\textwidth]{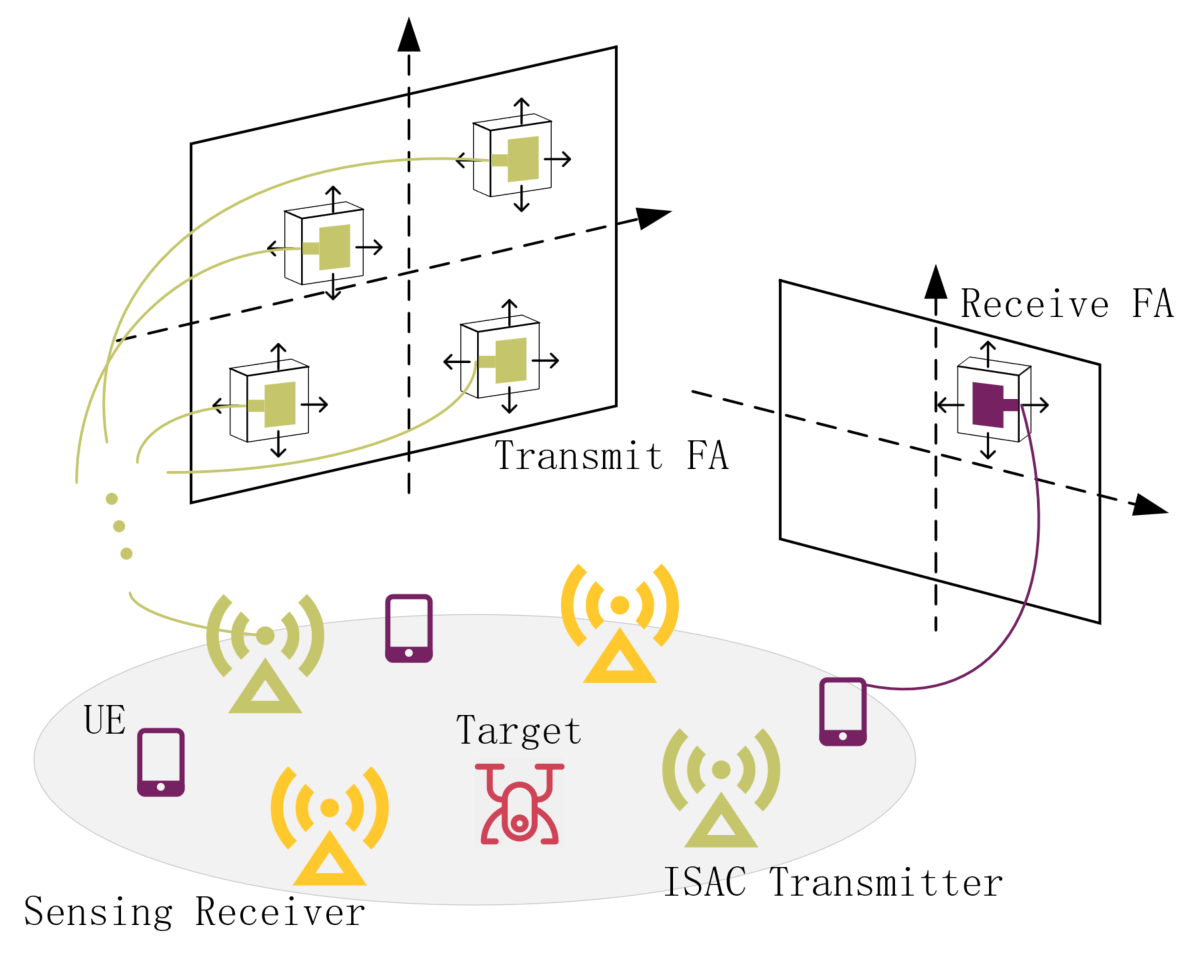} 
    \caption{DS-FAS-Assisted Multi-Static ISAC System.} 
    \label{System Model} 
\vspace{-15pt}
\end{figure}

We consider a DS-FAS-assisted multi-static ISAC system, where the set $\mathcal{M}_\textrm{t}\triangleq\left\{1,2,\dots,M_\textrm{t}\right\}$ of $M_\textrm{t}=|\mathcal{M}_\textrm{t}|$ ISAC transmitters jointly serve the set $\mathcal{K}\triangleq\left\{1,2,\dots,K\right\}$ of $K=|\mathcal{K}|$ UEs while detecting a potential target, and another set $\mathcal{M}_\textrm{r}\triangleq\left\{1,2,\dots,M_\textrm{r}\right\}$ of $M_\textrm{r}=|\mathcal{M}_\textrm{r}|$ APs serve as sensing receivers. To fully leverage the potential DoFs provided by FAS, we assume that each ISAC transmitter and sensing receiver is equipped with $N$ FAs, and each UE is equipped with a single FA, all of which can move within a local rectangular area as illustrated in Fig.~\ref{System Model}.
\vspace{-10pt}

\subsection{Channel Model}
Assuming that the moving regions of FAs on both sides are negligible compared to the signal propagation distance, for each channel path component, all FAs of each transmitter and receiver experience the same angle of departure/arrival (AoD/AoA) and the same amplitude of path gain, while only the phase coefficients differ among different FAs, which can be adjusted by changing FA positions~\cite{10318061},~\cite{10243545}.

\textit{1) Communication Channel Model:}

Considering a rectangular moving region for FAs at each AP, a 2D local coordinate system is established for ISAC transmitter $t$, whose antenna position matrix (APM) is denoted as $\mathbf{U}_t=\left[\mathbf{u}_{t,1},\mathbf{u}_{t,2},\dots,\mathbf{u}_{t,N}\right]\in\mathbb{R}^{2\times N}$, where $\mathbf{u}_{t,n}=[x_{t,n},y_{t,n}]^T\in\mathcal{C}^\textrm{t}$ denotes the $n$-th FA position, $\mathcal{C}^\textrm{t}\triangleq[-X^\textrm{t}/2,X^\textrm{t}/2]\times[-Y^\textrm{t}/2,Y^\textrm{t}/2]$ denotes the moving region of each FA at ISAC transmitters\footnote{It will be detailed in Section \ref{IV.B} that the positions of sensing receivers' FAs will not affect the performance of the considered system. Thus we only consider the optimization of APMs of ISAC transmitters and UEs.}. The number of propagation paths between ISAC transmitter $t$ and UE $k$ is defined as $L$. For the communication channel between ISAC transmitter $t$ and UE $k$, the field response vector for the $n$-th FA of ISAC transmitter $t$ is given by
\begin{equation}
    \mathbf{a}_{t,k,n}\left(\mathbf{u}_{t,n}\right) \triangleq [e^{-j\frac{2\pi}{\lambda} \boldsymbol{\varpi}^\textrm{t}_{t,k,1}\mathbf{u}_{t,n}},\cdots,e^{-j\frac{2\pi}{\lambda} \boldsymbol{\varpi}^\textrm{t}_{t,k,L}\mathbf{u}_{t,n}}]^T,
\end{equation}
where $\boldsymbol{\varpi}^\textrm{t}_{t,k,l} = [\textrm{sin}(\phi^\textrm{t}_{t,k,l})\textrm{cos}(\psi^\textrm{t}_{t,k,l}),\textrm{cos}(\phi^\textrm{t}_{t,k,l})]$, $\phi_{t,k,l}^\textrm{t}\in[0,\pi)$ and $\psi^\textrm{t}_{t,k,l}\in[0,\pi)$ represent the elevation and azimuth AoDs of the $l$-th propagation path between ISAC transmitter $t$ and UE $k$, respectively. Therefore, for the communication channel between ISAC transmitter $t$ and UE $k$, the field response matrix (FRM) of ISAC transmitter $t$ is 
\begin{equation}
    \mathbf{A}_{t,k}\left(\mathbf{U}_t\right) \triangleq [\mathbf{a}_{t,k,1}\left(\mathbf{u}_{t,1}\right),\cdots,\mathbf{a}_{t,k,N}\left(\mathbf{u}_{t,N}\right)].
\end{equation}

Similarly, from the UE side, a 2D local coordinate system is established for each UE $k$ and its APM 
is denoted as $\mathbf{v}_k=[x_{k},y_{k}]^T$ with $\mathbf{v}_k\in\mathcal{C}^\textrm{r}$, where $\mathcal{C}^\textrm{r}\triangleq[-X^\textrm{r}/2,X^\textrm{r}/2]\times[-Y^\textrm{r}/2,Y^\textrm{r}/2]$ denotes the moving region of each UE's FA. For the communication channel between ISAC transmitter $t$ and UE $k$, the FRM of UE $k$ is given by
\begin{equation}
    \mathbf{b}_{t,k}\left(\mathbf{v}_k\right) \triangleq [e^{-j\frac{2\pi}{\lambda}\boldsymbol{\varpi}^\textrm{r}_{t,k,1}\mathbf{v}_k},\cdots,e^{-j\frac{2\pi}{\lambda}\boldsymbol{\varpi}^\textrm{r}_{t,k,L}\mathbf{v}_k}]^T,
\end{equation}
where $\boldsymbol{\varpi}^\textrm{r}_{t,k,l} = [\textrm{sin}(\phi^\textrm{r}_{t,k,l})\textrm{cos}(\psi^\textrm{r}_{t,k,l}),\textrm{cos}(\phi^\textrm{r}_{t,k,l})]$, $\phi^\textrm{r}_{t,k,l}\in[0,\pi)$ and $ \psi^\textrm{r}_{t,k,l}\in[0,\pi)$ denote the elevation and azimuth AoAs of the $l$-th propagation path between ISAC transmitter $t$ and UE $k$, respectively.

Furthermore, the path response matrix (PRM) between ISAC transmitter $t$ and UE $k$ is defined as $\boldsymbol{\Sigma}_{t,k} \triangleq \textrm{diag}\left\{\upsilon_{t,k,1},\dots,\upsilon_{t,k,L}\right\}\in\mathbb{C}^{L\times L}$, where $\upsilon_{t,k,l}$ denotes the path response of the $l$-th propagation path between ISAC transmitter $t$ and UE $k$. As a result, the channel vector between ISAC transmitter $t$ and UE $k$ is given by 
\begin{equation}\label{h_tk}
    \mathbf{h}^H_{t,k}\left(\mathbf{U}_t,\mathbf{v}_k\right) = \mathbf{b}^H_{t,k}\left(\mathbf{v}_k\right)\boldsymbol{\Sigma}_{t,k}\mathbf{A}_{t,k}\left(\mathbf{U}_t\right).
\end{equation}

\textit{2) Sensing Channel Model:}

Consistent with previous research, line-of-sight (LoS) path is assumed to exist between APs and target while non-line-of-sight (NLoS) paths are neglected. Building on this, the steering vector between ISAC transmitter $t$ and the target is given by
\begin{equation}
    \mathbf{a}_{t,0}\left(\mathbf{U}_t\right) \triangleq [e^{-j\frac{2\pi}{\lambda} \boldsymbol{\varpi}^\textrm{t}_{t,0}\mathbf{u}_{t,1}},\cdots,e^{-j\frac{2\pi}{\lambda} \boldsymbol{\varpi}^\textrm{t}_{t,0}\mathbf{u}_{t,N}}]^T,
\end{equation}
where $\boldsymbol{\varpi}^\textrm{t}_{t,0} = [\textrm{sin}(\phi^\textrm{t}_{t,0})\textrm{cos}(\psi^\textrm{t}_{t,0}),\textrm{cos}(\phi^\textrm{t}_{t,0})]$, $\phi^\textrm{t}_{t,0}\in[0,\pi)$ and $ \psi^\textrm{t}_{t,0}\in[0,\pi)$ represent the elevation and azimuth AoDs of the LoS path between ISAC transmitter $t$ and the target.

As for the sensing receiver side, since the sensing receivers' FAs will not affect the performance as will be certified later, we simply denote the steering vector between sensing receiver $r$ and the target as $\mathbf{b}_{r,0} \triangleq \left[b_{r,0,1},\dots,b_{r,0,N}\right]^T\in\mathbb{C}^{N\times 1}$ without loss of optimality, where $|b_{r,0,n}|=1,1\leq n\leq N$.

Based on the principle of wireless sensing, the round-trip channel from ISAC transmitter $t$ to the target and back to sensing receiver $r$ is given by
\begin{equation}
    \mathbf{H}_{t,r}\left(\mathbf{U}_t\right) = \alpha_{t,r}\mathbf{G}_{t,r}\left(\mathbf{U}_t\right) \in\mathbb{C}^{N\times N},
\end{equation}
where $\mathbf{G}_{t,r}\left(\mathbf{U}_t\right)\triangleq\sqrt{\beta_{t,r}}\mathbf{b}_{r,0}\mathbf{a}^H_{t,0}\left(\mathbf{U}_t\right)$, $\alpha_{t,r}\sim\mathcal{CN}(0,\sigma_{t,r}^2)$ is the target radar cross section (RCS) corresponding to the round-trip propagation path, $\sigma_{t,r}^2$ denotes the variance of the target RCS, $\beta_{t,r}$ represents the coefficient characterizing the path loss~\cite{10494224}.

\subsection{ISAC Signal Model}\label{ISAC Signal}

In the considered DS-FAS-assisted multi-static ISAC system, the data symbol intended for UE $k$ is denoted as $c_k$, which is of unit power, i.e., $\mathbb{E}\left\{|c_k|^2\right\}=1$. After being digitally beamformed by $\mathbf{w}_{t,k}\in\mathbb{C}^{N\times 1}$, $c_k$ is transmitted from ISAC transmitter $t$, which is utilized for both S\&C. To fully reap the DoFs provided by multiple FAs at the transmitter side, an additional dedicated sensing signal $\mathbf{s}_{t,0}$ is transmitted by each ISAC transmitter $t$. Therefore, at time instance $\tau$, the ISAC signal transmitted from ISAC transmitter $t$ is given by
\begin{equation}
    \mathbf{x}_t[\tau] 
    =  \sum\nolimits_{k\in\mathcal{K}}\mathbf{w}_{t,k}c_k[\tau] + \mathbf{s}_{t,0}[\tau].
\end{equation}
It should be noted that different from the data symbol for each UE that is transmitted with a single beam, the dedicated sensing signal can be transmitted with multiple beams~\cite{9124713}, i.e., $\textrm{tr}\left(\mathbf{R}_0\right)\geq 1$, where $\mathbf{R}_0\triangleq\mathbb{E}\left\{\mathbf{s}_0[\tau]\mathbf{s}_0[\tau]^H\right\}$, $\mathbf{s}_0[\tau]\triangleq[\mathbf{s}^H_{1,0}[\tau],\dots,\mathbf{s}^H_{M_\textrm{t},0}[\tau]]^H\in\mathbb{C}^{NM_\textrm{t}\times 1}$.

\textit{1) Signal Received at UE}

The received signal at UE $k$ at time instance $\tau$ can be expressed as 
\begin{equation}
\begin{aligned}
    y_k[\tau] &= \sum\nolimits_{t=1}^{M_\textrm{t}} \mathbf{h}_{t,k}^H\mathbf{x}_t[\tau] + n_k[\tau]=\mathbf{h}_k^H\mathbf{x}[\tau] + n_k[\tau],
\end{aligned}
\end{equation}
where $\mathbf{h}_k\triangleq[\mathbf{h}_{1,k}^H,\dots,\mathbf{h}^H_{M_\textrm{t},k}]^H\in\mathbb{C}^{NM_\textrm{t}\times 1}$, $\mathbf{x}[\tau]\triangleq[\mathbf{x}^H_1[\tau],\dots,\mathbf{x}^H_{M_\textrm{t}}[\tau]]^H\in\mathbb{C}^{NM_\textrm{t}\times 1}$, and $n_k[\tau]\sim\mathcal{CN}\left(0,\sigma_k^2\right)$ denotes the additive white Gaussian noise (AWGN) received at UE $k$ at time instance $\tau$. Then, the SINR of UE $k$ is given as 
\begin{equation}\label{SINR}
		\textrm{SINR}_k\left(\mathbf{w},\mathbf{s}_0,\mathbf{u},\mathbf{v}_k\right) =\frac{\left|\mathbf{h}^H_{k}\left(\mathbf{u},\mathbf{v}_k\right)\mathbf{w}_k\right|^2}{I_k\left(\mathbf{w},\mathbf{s}_0,\mathbf{u},\mathbf{v}_k\right)+\sigma_k^2},\ k\in\mathcal{K}
	\end{equation}
where $I_k\left(\mathbf{w},\mathbf{s}_0,\mathbf{u},\mathbf{v}_k\right)\triangleq\sum\limits_{j\neq k}\left|\mathbf{h}^H_{k}\left(\mathbf{u},\mathbf{v}_k\right)\mathbf{w}_j\right|^2 + \mathbf{h}^H_{k}\left(\mathbf{u},\mathbf{v}_k\right)$ $\mathbf{R}_0\mathbf{h}_{k}\left(\mathbf{u},\mathbf{v}_k\right)$, $\mathbf{w}_k\triangleq[\mathbf{w}_{1,k}^H,\dots,\mathbf{w}_{M_\textrm{t},k}^H]^H\in\mathbb{C}^{NM_\textrm{t}\times 1}$, $k\in\mathcal{K}$.

\textit{2) Signal Received at Sensing Receiver}

Then, the received signal at sensing receiver $r$ at time instance $\tau$ can be expressed as 
\begin{equation}
    \mathbf{y}_r[\tau] = \sum\nolimits_{t=1}^{M_\textrm{t}} \mathbf{H}_{t,r}\mathbf{x}_t[\tau] + \mathbf{n}_r[\tau]=\mathbf{G}_r\mathbf{X}[\tau]\boldsymbol{\alpha}_r + \mathbf{n}_r[\tau],
\end{equation}
where $\boldsymbol{\alpha}_r\triangleq[\alpha_{1,r},\cdots,\alpha_{M_\textrm{t},r}]^T\in\mathbb{C}^{M_\textrm{t}\times 1}$, $\mathbf{X}[\tau]\triangleq\textrm{blkdiag}\left(\right.$ $\left.\mathbf{x}_1[\tau],\cdots,\mathbf{x}_{M_\textrm{t}}[\tau]\right)\in\mathbb{C}^{NM_\textrm{t}\times M_\textrm{t}}$, $\mathbf{G}_r\triangleq [\mathbf{G}_{1,r},\cdots,\mathbf{G}_{M_\textrm{t},r}]\in\mathbb{C}^{N\times NM_\textrm{t}}$, and $\mathbf{n}_r[\tau]\sim\mathcal{CN}\left(\mathbf{0},\sigma_r^2\mathbf{I}_N\right)$ denotes the AWGN received at sensing receiver $r$ at time instance $\tau$.

\section{GLRT Detector for Multi-Static Sensing}\label{GLRT}
Typically, the detection probability is a widely-used metric to assess the detection performance\cite{10494224}, \cite{10605793}. In this section, the GLRT detector is applied for target detection, based on which the generalized likelihood ratio (GLR) is first derived, and the detection probability is finally obtained as a function of the optimization variables.

\vspace{-2pt}
\subsection{Binary Hypothesis and GLR Derivation}
In this paper, received signals in $T$ time instances are utilized for detecting target at a particular location. The concatenated received signal over all $M_\textrm{r}$ sensing receivers at time instance $\tau$ can be expressed as
\begin{equation}
    \mathbf{y}[\tau] = [\mathbf{y}_r^H[\tau],\cdots,\mathbf{y}_{M_\textrm{r}}^H[\tau]]^H=\mathbf{G}\left(\mathbf{I}_{M_\textrm{r}}\otimes\mathbf{X}[\tau]\right)\boldsymbol{\alpha} + \mathbf{n}[\tau],
\end{equation}
where $\mathbf{G} \triangleq \textrm{blkdiag}\left(\mathbf{G}_1,\cdots,\mathbf{G}_{M_\textrm{r}}\right)\in\mathbb{C}^{NM_\textrm{r}\times NM_\textrm{r}M_\textrm{t}}$, $\boldsymbol{\alpha}\triangleq[\boldsymbol{\alpha}^H_1,\cdots,\boldsymbol{\alpha}^H_{M_\textrm{r}}]^H\in\mathbb{C}^{M_\textrm{r}M_\textrm{t}\times 1}$, $\mathbf{n}[\tau]\triangleq\big[\mathbf{n}_1^H[\tau],\dots,\big.$ $\big.\mathbf{n}_{M_\textrm{r}}^H[\tau]\big]^H\mathbb{C}^{NM_\textrm{r}\times 1}$. Then, the concatenated received signal over total $T$ time instances can be expressed as
\begin{equation}
\begin{aligned}
    \mathbf{y} &= [\mathbf{y}^H[1],\cdots,\mathbf{y}^H[T]]^H= \tilde{\mathbf{G}}\boldsymbol{\alpha} + \mathbf{n},
\end{aligned}
\end{equation}
where $\tilde{\mathbf{G}}\triangleq\big[\tilde{\mathbf{G}}^H[1],\cdots,\tilde{\mathbf{G}}^H[T]\big]^H$, $\tilde{\mathbf{G}}[\tau]\triangleq\mathbf{G}\left(\mathbf{I}_{M_\textrm{r}}\otimes\mathbf{X}[\tau]\right)$ $\in\mathbb{C}^{NM_\textrm{r}\times M_\textrm{r}M_\textrm{t}}$, $\mathbf{n}\triangleq\left[\mathbf{n}^H[1],\right.$ $\left.\dots,\mathbf{n}^H[T]\right]^H\in\mathbb{C}^{TNM_\textrm{r}\times 1}$. 

Without diving into a specific statistical model for the RCS, we model $\boldsymbol{\alpha}$ as unknown but deterministic parameter as in~\cite{10639146},~\cite{10605793}. Therefore, whether $\boldsymbol{\alpha}=\mathbf{0}$ or not represents whether the target of interest exists or not. Then, the binary hypothesis for GLRT detector can be written as
\begin{equation}\label{B-H}
    \begin{cases}
        \mathcal{H}_1:\  \mathbf{y} = {\tilde{\mathbf{G}}}\boldsymbol{\alpha} + \mathbf{n}, \\
        \mathcal{H}_0:\ \mathbf{y} = \mathbf{n}.
    \end{cases}
\end{equation}

\textit{Proposition 1:} Given the binary hypothesis \eqref{B-H}, the GLR of the GLRT detector is given as
 \begin{equation}
        L_\textrm{G}(\mathbf{y}) = \exp\big(\frac{1}{\sigma^2}\mathbf{y}^H\tilde{\mathbf{G}}(\tilde{\mathbf{G}}^H\tilde{\mathbf{G}})^{-1}\tilde{\mathbf{G}}^H\mathbf{y}\big)
\end{equation}

\textit{Proof:} See Appendix \ref{APP1}.

Then, the decision can be made by GLRT detector as
\begin{equation}
    \hat{\mathcal{H}}=\begin{cases}
        \mathcal{H}_1:\  L_\textrm{G}(\mathbf{y}) \geq \gamma_0, \\
        \mathcal{H}_0:\ L_\textrm{G}(\mathbf{y}) < \gamma_0.
    \end{cases}
\end{equation}
where $\gamma_0$ is the detection threshold. 

\vspace{-2pt}
\subsection{Detection Probability Derivation}

To derive the detection probability, the probability density function (PDF) of the GLR is needed. 

\textit{Proposition 2}: Given $\boldsymbol{\alpha}$, the asymptotic distribution of $\textrm{ln}\left(L_\textrm{G}(\mathbf{y})\right)=\frac{1}{\sigma^2}\mathbf{y}^H\tilde{\mathbf{G}}$ $(\tilde{\mathbf{G}}^H\tilde{\mathbf{G}})^{-1}\tilde{\mathbf{G}}^H\mathbf{y}$ is given as
\begin{equation}
    \textrm{ln}\left(L_\textrm{G}(\mathbf{y})\right)\sim\begin{cases}
        \chi_{2M_\textrm{r}M_\textrm{t}}^2,&\mathcal{H}_0, \\
        \tilde{\chi}_{2M_\textrm{r}M_\textrm{t}}^2\left(\omega\right),&\mathcal{H}_1,
    \end{cases}
\end{equation}
where $\chi_{2M_\textrm{r}M_\textrm{t}}^2$ and $\tilde{\chi}_{2M_\textrm{r}M_\textrm{t}}^2\left(\omega\right)$ are central and non-central chi-squared distributions with $2M_\textrm{r}M_\textrm{t}$ DoFs,  respectively, and $\omega = \frac{\boldsymbol{\alpha}^H\tilde{\mathbf{G}}^H\tilde{\mathbf{G}}\boldsymbol{\alpha}}{\sigma^2}$.

\textit{Proof:} See Appendix \ref{APP2}.

Given the PDF of GLR, the false-alarm probability can be calculated first as $P_\textrm{FA}\left(\gamma_0\right)=P\left(L_\textrm{G}\left(\mathbf{y}\right)\geq \gamma_0;\boldsymbol{\alpha},\mathcal{H}_0\right)$. Then, we resort to the Neyman-Pearson test with the given desired false-alarm probability $\hat{P}_\textrm{FA}$, i.e., letting $P_\textrm{FA}\left(\gamma_0\right)=1-F_{\chi_{2M_\textrm{r}M_\textrm{t}}^2}\left(\textrm{ln}(\gamma_0)\right)=\hat{P}_\textrm{FA}$, where $F_{\chi_{2M_\textrm{r}M_\textrm{t}}^2}(\cdot)$ denotes the cumulative distribution function of central chi-squared distribution. Then the detection threshold can be calculated as $\gamma_0=\textrm{exp}(F_{\chi_{2M_\textrm{r}M_\textrm{t}}^2}^{-1}(1-\hat{P}_\textrm{FA}))$. Based on the calculated $\gamma_0$, the detection probability can be finally derived as $P_\textrm{D}\left(\omega\right)=1-F_{\tilde{\chi}_{2M_\textrm{r}M_\textrm{t}}^2\left(\omega\right)}\left(\textrm{ln}(\gamma_0)\right)$, where $F_{\tilde{\chi}_{2M_\textrm{r}M_\textrm{t}}^2\left(\cdot\right)}\left(\cdot\right)$ denotes the cumulative distribution function of non-central chi-squared distribution, $\omega$ is a function of $\left\{\mathbf{w}\right\}$, $\left\{\mathbf{s}_{0}\right\}$, and $\left\{\mathbf{u}\right\}$.

\section{Problem Formulation and Reformulation}\label{IV}

\subsection{Problem Formulation}
To maximize the sensing performance while ensuring the communication performance of each UE, the optimization problem can be formulated as
\begin{equation}
\begin{aligned}
    \left(\mathscr{P}_1\right)&:  \mathop{\textrm{max}}\limits_{\mathbf{w},\mathbf{s}_0,\mathbf{u},\mathbf{v}}P_\textrm{D}\left(\mathbf{w},\mathbf{s}_0,\mathbf{u}\right)\\
    \textrm{s.t.}\enspace&\textrm{C1}:\textrm{SINR}_k\left(\mathbf{w},\mathbf{s}_0,\mathbf{u},\mathbf{v}\right)\geq \gamma_k,\ \forall k\in\mathcal{K},\\
    &\textrm{C2}:p_t\left(\mathbf{w},\mathbf{s}_0\right) \leq P_t,\ \forall t\in\mathcal{M}_\textrm{t},\\
    &\textrm{C3}: \mathbf{u}_{t,n}\in\mathcal{C}^\textrm{t},\ \forall t\in\mathcal{M}_\textrm{t},1\leq n\leq N,\\
    &\textrm{C4}:\mathbf{v}_k\in\mathcal{C}^\textrm{r},\ \forall k\in\mathcal{K},\\
    &\textrm{C5}: \left\Vert\mathbf{u}_{t,i}-\mathbf{u}_{t,j}\right\Vert_2 \geq \lambda/2, \forall t\in\mathcal{M}_\textrm{t}, 1\leq i,j\leq N, i\neq j.
\end{aligned}
\end{equation}
In the above problem, constraint C1 shows the QoS requirement of each UE $k$ with $\gamma_k$ as the minimum SINR requirement,
constraint C2 shows the power budget $P_t$ of each ISAC transmitter $t$. The power consumption of each ISAC transmitter $p_t\left(\mathbf{w},\mathbf{s}_0\right)$ can be expressed as
\begin{equation}
    p_t\left(\mathbf{w},\mathbf{s}_0\right) = \sum\nolimits_{k\in\mathcal{K}}\mathbf{w}_{t,k}^H\mathbf{w}_{t,k} + \mathbb{E}\left\{\mathbf{s}_{t,0}^H[\tau]\mathbf{s}_{t,0}[\tau]\right\}.
\end{equation}
Constraints C3 and C4 denote the moving region of ISAC transmitters' FAs and UEs' FAs, respectively. To mitigate the coupling effect between different elements of a FA array, a separation distance $\lambda/2$ between adjacent elements has to be ensured, which is denoted as constraint C5.

\subsection{Problem Reformulation}\label{IV.B}

It can be observed that the detection probability $P_\textrm{D}$ is a monotonically increasing function with respect to $\omega$. Therefore, maximizing $P_\textrm{D}$ is equivalent to maximizing $\omega$. Furthermore, since $\omega=\frac{\boldsymbol{\alpha}^H\tilde{\mathbf{G}}^H\tilde{\mathbf{G}}\boldsymbol{\alpha}}{\sigma^2}=\frac{\textrm{tr}\left(\boldsymbol{\alpha}\boldsymbol{\alpha}^H\tilde{\mathbf{G}}^H\tilde{\mathbf{G}}\right)}{\sigma^2}$, we denote the objective function as $\omega = \textrm{tr}(\boldsymbol{\Psi}\tilde{\mathbf{G}}^H\tilde{\mathbf{G}})$, where $\boldsymbol{\Psi}=\boldsymbol{\alpha}\boldsymbol{\alpha}^H/\sigma^2$.

To derive a more tractable form of problem with objective function $\omega$, we define the covariance matrix of the transmitted signal $\mathbf{x}[\tau]$ as
\begin{equation}\label{R}
    \mathbf{R}\left(\mathbf{w},\mathbf{s}_0\right)\triangleq\mathbb{E}\left\{\mathbf{x}[\tau]\mathbf{x}^H[\tau]\right\}=\sum\nolimits_{k\in\mathcal{K}}\mathbf{w}_k\mathbf{w}_k^H+\mathbf{R}_0.
\end{equation}
Then $\omega$ can be reformulated as
\begin{equation}\label{OBJ}
    \begin{aligned}
        \omega=TN\sum\nolimits_{i=1}^{M_\textrm{t}}\sum\nolimits_{j=1}^{M_\textrm{t}}&\Big(\sum\nolimits_{r=1}^{M_\textrm{r}}\left[\boldsymbol{\Psi}_{r}\right]_{i,j}\sqrt{\beta_{i,r}\beta_{j,r}}\Big)\\
        &\quad\cdot\mathbf{a}^H_{i,0}\left(\mathbf{U}_i\right)\mathbf{E}_i\mathbf{R}\mathbf{E}_j^H\mathbf{a}_{j,0}\left(\mathbf{U}_j\right)
    \end{aligned}
\end{equation}
where $\boldsymbol{\Psi}_r\in\mathbb{C}^{M_\textrm{t}\times M_\textrm{t}}$ is the $r$-th diagonal sub-block of $\boldsymbol{\Psi}$. $\mathbf{E}_i\in\mathbb{C}^{N\times NM_\textrm{t}}, 1\leq i\leq M_\textrm{t}$ is defined as a block selection matrix composed of $M_\textrm{t}$ horizontally arranged square matrices, where only the $i$-th square matrix is set as identity matrix, while others are all set as zero matrices, i.e.,
\begin{equation}\label{E}
    \mathbf{E}_i\triangleq[\mathbf{0},\dots,\mathbf{I}_N,\dots,\mathbf{0}],
\end{equation}
$\mathbf{E}_i\mathbf{A}\mathbf{E}_j^H$ is utilized to extract the sub-block at the $i$-th row, $j$-th column of matrix $\mathbf{A}\in\mathbb{C}^{NM_\textrm{t}\times NM_\textrm{t}}$, and $[\mathbf{A}]_{i,j}$ denotes the element at the $i$-th row $j$-th column of matrix $\mathbf{A}$. The derivation of \eqref{OBJ} is given in Appendix \ref{APP3}.

\textit{Remark 1: It can be observed from \eqref{OBJ} that the objective function is independent of $\mathbf{b}_{r,0}$. Therefore, the positions of sensing receivers' FAs will not affect the performance of the considered system. However, when more than one target is considered, the design of sensing receivers' FA positions is required, which is left for future investigation.}

Based on definitions \eqref{R} and \eqref{E}, the power constraint C2 can be reformulated as
\begin{equation}\label{C2'}
    \textrm{C2}':\ p_t(\mathbf{w},\mathbf{s}_0)=\textrm{tr}\left(\mathbf{E}_t\mathbf{R}\left(\mathbf{w},\mathbf{s}_0\right)\mathbf{E}_t^H\right)\leq P_t,\ \forall t\in\mathcal{M}_\textrm{t},
\end{equation}
and the optimization problem $\left(\mathscr{P}_1\right)$ can be equivalently expressed as
\begin{equation}
    \left(\mathscr{P}_2\right):\mathop{\textrm{max}}\limits_{\mathbf{w},\mathbf{s}_0,\mathbf{u},\mathbf{v}}\omega\left(\mathbf{w},\mathbf{s}_0,\mathbf{u}\right)\quad\textrm{s.t.}\enspace\textrm{C1},\textrm{C2}',\textrm{C3},\textrm{C4},\textrm{C5}.
\end{equation}

To avoid the computationally demanding task of finding an initial feasible point $\{\mathbf{w}^{(0)},\mathbf{s}^{(0)}_0,\mathbf{u}^{(0)},\mathbf{v}^{(0)}\}$ that satisfies the potentially stringent QoS requirements of UEs, we resort to the penalty-based method to relax constraint $\textrm{C1}$. To prevent introducing additional non-convexity, rather than directly adding the slack variable to the original form, we first reformulate $\textrm{C1}$ as \eqref{C1'} shown at the bottom of next page.
\begin{figure*}[hb]
\vspace{-2pt}
	\centering
	\hrulefill
	\vspace*{0pt} 
	\begin{equation}\label{C1'}
		\textrm{C1}':\zeta_k\left(\mathbf{w},\mathbf{s}_0,\mathbf{u},\mathbf{v}_k\right) \triangleq\left|\mathbf{h}^H_{k}\left(\mathbf{u},\mathbf{v}_k\right)\mathbf{w}_k\right|^2-\gamma_k\Big({\sum_{j\neq k}\left|\mathbf{h}^H_{k}\left(\mathbf{u},\mathbf{v}_k\right)\mathbf{w}_j\right|^2 + \mathbf{h}^H_{k}\left(\mathbf{u},\mathbf{v}_k\right)\mathbf{R}_0\mathbf{h}_{k}\left(\mathbf{u},\mathbf{v}_k\right)+\sigma_k^2}\Big)\geq 0,\ k\in\mathcal{K}
	\end{equation}
\end{figure*}
Subsequently, an auxiliary slack variable $\nu>0$ is introduced into $\textrm{C1}'$ and incorporated into the objective function as a penalty term. Overall, problem $\left(\mathscr{P}_2\right)$ can be reformulated as
\begin{equation}
\begin{aligned}
    \left(\mathscr{P}_3\right): &\mathop{\textrm{max}}\limits_{\mathbf{w},\mathbf{s}_0,\mathbf{u},\mathbf{v},\nu}\ \rho\left(\mathbf{w},\mathbf{s}_0,\mathbf{u},\nu\right)\triangleq\omega\left(\mathbf{w},\mathbf{s}_0,\mathbf{u}\right) - \eta\cdot\nu\\
    \textrm{s.t.}\enspace&\textrm{C1}'':\zeta_k\left(\mathbf{w},\mathbf{s}_0,\mathbf{u},\mathbf{v}_k\right)+\nu\geq 0,\ \forall k\in\mathcal{K}\\
    &\textrm{C2}',\textrm{C3},\textrm{C4},\textrm{C5},\textrm{C6}:\nu\geq 0,
\end{aligned}
\end{equation}
where $\eta>0$ denotes a sufficiently large penalty factor used to penalize any violation of the original constraint $\textrm{C1}$. If the original QoS requirements are feasible, the penalty term $\nu$ will vanish upon convergence, otherwise, a non-zero $\nu$ indicates the infeasibility of the original problem. It can be observed that $\nu$ is introduced as a linear additive term in both objective function and the constraint, and it is not coupled with any other optimization variables.

To gain more insights into the impact of transmit FAs on the sensing performance, we consider the sensing-only scenario:

\textit{Proposition 3:} For the sensing-only scenario with optimization problem
\begin{equation}\label{P3}
    \left(\mathscr{P}_4\right): \mathop{\textrm{max}}\limits_{\mathbf{R},\mathbf{u}}\ \omega\left(\mathbf{R},\mathbf{u}\right)\enspace\textrm{s.t.}\enspace\textrm{C2}',\textrm{C3},\textrm{C5},\textrm{C7}:\mathbf{R}\succeq\mathbf{0},
\end{equation}
the optimal performance can be achieved by only optimizing the transmit beamforming (denoted by $\mathbf{R}$) for any given transmit FA positions $\mathbf{u}$ satisfying constraint C3 and C5. 

\textit{Proof:} See Appendix \ref{APP4}.

\textit{Remark 2: Based on Proposition 3, although the design of transmit FA positions will not affect the performance in the sensing-only scenario, due to the conflict between the S\&C functions for power and spatial resources (denoted by $\mathbf{w}$ and $\mathbf{s}_0$), the design of $\mathbf{u}$ can help satisfy constraint $\textrm{C1}$ in the ISAC scenario, and thereby release more DoFs of $\mathbf{w}$ and $\mathbf{s}_0$ to further enhance the sensing performance indicator $\omega$, which we term as "indirect effect". It is apparent that the optimization of positions of UEs' FAs also has a similar indirect effect on the sensing performance in the considered ISAC scenario. The principle of this mechanism will be utilized to design a more efficient algorithm in Section \ref{V.C}.}

\vspace{0pt}
\section{Proposed Solution for DS-FAS-Assisted Multi-Static ISAC}
In Section \ref{IV}, the optimization problem has been reformulated as $\left(\mathscr{P}_3\right)$, where both the objective function and feasible set are non-concave/non-convex. Furthermore, the parameters to be optimized are highly coupled, which makes $\left(\mathscr{P}_3\right)$ NP-hard. To address this issue, an AO-based scheme is proposed to decompose problem $\left(\mathscr{P}_3\right)$ into three subproblems. In each iteration, each subproblem updates a group of variables while keeping others fixed. 

For ease of notation, we drop the fixed variables in the description of functions. For instance, for an arbitrary function $f(x,y,z)$ of variables $x,y,z$, we use $f(x)$ to simplify the expression of $f\left(x,y^{(\kappa-1)},z^{(\kappa-1)}\right)$, where $(\cdot)^{(\kappa-1)}$ denotes the variable $(\cdot)$ obtained from the previous corresponding sub-problem.

\vspace{0pt}
\subsection{Optimizing $(\mathbf{w},\mathbf{s}_0,\nu)$ with given $(\mathbf{u}^{(\kappa-1)},\mathbf{v}^{(\kappa-1)})$}

This subproblem can be written as
\begin{equation}
\begin{aligned}
    \left(\mathscr{P}_{3.1}\right): &\mathop{\textrm{max}}\limits_{\mathbf{w},\mathbf{s}_0,\nu}\  \rho\left(\mathbf{w},\mathbf{s}_0,\nu\right)=\omega\left(\mathbf{w},\mathbf{s}_0\right)-\eta\cdot\nu\\
    \textrm{s.t.}\enspace&\textrm{C1}'':\zeta_k\left(\mathbf{w},\mathbf{s}_0\right)+\nu\geq 0,\ \forall k\in\mathcal{K},\textrm{C2}',\textrm{C6}.
\end{aligned}
\end{equation}
Resorting to the SDR-based approach, we define $\mathbf{R}_k\triangleq\mathbf{w}_k\mathbf{w}_k^H\succeq\mathbf{0}$ with $\textrm{rank}\left\{\mathbf{R}_k\right\}=1,\forall k\in\mathcal{K}$. Therefore, constraint $\textrm{C1}''$ can be finally reformulated as $\textrm{C1}'''$:
\begin{equation}\label{SINR1}
    	\underbrace{\mathbf{h}^H_{k}\left(\mathbf{u},\mathbf{v}_k\right)\left[(1+\gamma_k)\mathbf{R}_k-\gamma_k\mathbf{R}\right]\mathbf{h}_{k}\left(\mathbf{u},\mathbf{v}_k\right) - \gamma_k\sigma_k^2}_{\triangleq\tilde{\zeta}_k\left(\mathbf{R}_k,\mathbf{R},\mathbf{u},\mathbf{v}_k\right)}+\nu \geq 0,
\end{equation}
which depends on $\mathbf{R}_k$ and $\mathbf{R}$.
Since the sensing performance indicator $\omega$ and the power consumption $p_t$ can also be expressed as functions of $\mathbf{R}$ according to \eqref{OBJ} and \eqref{C2'}, we can equivalently transform problem $(\mathscr{P}_{3.1})$ as
\begin{equation}
\begin{aligned}
    \left(\mathscr{P}_{3.1}'\right): &\mathop{\textrm{max}}\limits_{\mathbf{R}_k,\mathbf{R},\nu}\ \rho\left(\mathbf{R},\nu\right)=\omega\left(\mathbf{R}\right)-\eta\cdot\nu\\
    \textrm{s.t.}\enspace&\textrm{C1}''':\tilde{\zeta}_k\left(\mathbf{R}_k,\mathbf{R}\right)+\nu\geq 0,\ \forall k\in\mathcal{K},\textrm{C2}', \textrm{C6},\\
    &\textrm{C8}:\mathbf{R}\succeq \mathbf{0},\mathbf{R}_k\succeq \mathbf{0},\ \forall k\in\mathcal{K},\\
    &\textrm{C9}:\textrm{rank}\left(\mathbf{R}_k\right)=1\ \forall k\in\mathcal{K}.
\end{aligned}
\end{equation}
Since $\rho\left(\mathbf{R},\nu\right)$, $\tilde{\zeta}_k\left(\mathbf{R}_k,\mathbf{R}\right)$ and $p_t\left(\mathbf{R}\right)$ are all affine functions with respect to $\mathbf{R}_k$ and $\mathbf{R}$, and the only non-convex constraint is the rank constraint $\textrm{C9}$. Therefore, the rank constraint is discarded and problem $\left(\mathscr{P}_{3.1}'\right)$ can be efficiently solved via off-the-shelf toolboxes, such as CVX.

\textit{Remark 3: It should be noted that although the discarding of rank constraint may lead to solution $\hat{\mathbf{R}}_k^{(\kappa-1)},\hat{\mathbf{R}}^{(\kappa-1)}$ with $\textrm{rank}(\hat{\mathbf{R}}_k^{(\kappa-1)})>1$, we can always construct another set of $\mathbf{R}_k^{(\kappa-1)},\mathbf{R}^{(\kappa-1)}$ with $\textrm{rank}(\mathbf{R}_k^{(\kappa-1)})=1$ that can achieve the same performance as that of $\hat{\mathbf{R}}_k^{(\kappa-1)},\hat{\mathbf{R}}^{(\kappa-1)}$. Therefore, the discard of rank constraint will not affect the optimality of problem $\left(\mathscr{P}_{3.1}'\right)$. The way to construct $\mathbf{R}_k^{(\kappa-1)},\mathbf{R}^{(\kappa-1)}$ with given $\hat{\mathbf{R}}_k^{(\kappa-1)},\hat{\mathbf{R}}^{(\kappa-1)}$, which is omitted due to space limitations, is available in many previous papers such as~\cite{9124713},~\cite{10605793}. Finally, the corresponding $\mathbf{w}^{(\kappa-1)},\mathbf{s}_0^{(\kappa-1)}$ can be obtained through the singular value decomposition (SVD).}

\vspace{0pt}
\subsection{Optimizing $(\mathbf{u},\nu)$ with given $(\mathbf{w}^{(\kappa-1)},\mathbf{s}_0^{(\kappa-1)},\mathbf{v}^{(\kappa-1)})$}\label{V.B}

This subproblem can be given as
\begin{equation}
\begin{aligned}
    \left(\mathscr{P}_{3.2}\right): &\mathop{\textrm{max}}\limits_{\mathbf{u},\nu}\ \rho\left(\mathbf{u},\nu\right)=\omega\left(\mathbf{u}\right)-\eta\cdot\nu\\
    \textrm{s.t.}\enspace&\textrm{C1}''':\tilde{\zeta}_k\left(\mathbf{u}\right)+\nu\geq 0,\ \forall k\in\mathcal{K},\textrm{C3},\textrm{C5},\textrm{C6},
\end{aligned}
\end{equation}
where both the non-central parameter $\omega(\mathbf{u})$ and constraint $\textrm{C1}'''$ are non-concave/non-convex with respect to $\mathbf{u}$. Due to the tight coupling between the positions of different FA elements, we propose to optimize the position of each FA element $\mathbf{u}_{t,n}$, while keeping positions of other FA elements fixed. For the FA element $\mathbf{u}_{t,n}$, we separately process constraint $\textrm{C1}'''$ and $\omega(\mathbf{u})$ as follows:

\textit{1) Constraint $\textrm{C1}'''$ for given $\left\{\mathbf{u}_{t',n'},(t',n')\neq(t,n)\right\}$}

Based on \eqref{SINR1}, $\tilde{\zeta}_k\left(\mathbf{u}_{t,n}\right)$ can be rewritten as 
\begin{equation}\label{zeta}
\begin{aligned}    \tilde{\zeta}_k\left(\mathbf{u}_{t,n}\right)=&\mathbf{a}^H_{t,k,n}\left(\mathbf{u}_{t,n}\right)\mathbf{P}_{t,k,n}\mathbf{a}_{t,k,n}\left(\mathbf{u}_{t,n}\right)\\
&\quad\quad+ 2\mathscr{R}\left\{\mathbf{a}^H_{t,k,n}\left(\mathbf{u}_{t,n}\right)\mathbf{q}_{t,k,n}\right\} + \epsilon_{t,k,n},
\end{aligned}
\end{equation}
where $\epsilon_{t,k,n}$ is the term independent of $\mathbf{u}_{t,n}$, $\mathbf{P}_{t,k,n}$ and $\mathbf{q}_{t,k,n}$ are defined as
\begin{equation}\label{P_tkn}
    \mathbf{P}_{t,k,n} = [\mathbf{E}_t\tilde{\mathbf{R}}^{(\kappa-1)}_k\mathbf{E}_t^H]_{n,n}\boldsymbol{\Sigma}_{t,k}^H\mathbf{b}_{t,k}\mathbf{b}^H_{t,k}\boldsymbol{\Sigma}_{t,k}\in\mathbb{C}^{L\times L}
\end{equation}
and
\begin{equation}
    \begin{aligned}
        \mathbf{q}_{t,k,n}=&\sum\nolimits_{(t',n')\neq(t,n)}[\mathbf{E}_{t'}\tilde{\mathbf{R}}^{(\kappa-1)}_k\mathbf{E}_t^H]_{n',n}\\
        &\qquad\qquad\boldsymbol{\Sigma}_{t,k}^H\mathbf{b}_{t,k}\mathbf{b}^H_{t',k}\boldsymbol{\Sigma}_{t',k}\mathbf{a}_{t',k,n'}\left(\mathbf{u}_{t',n'}\right)\in\mathbb{C}^{L\times 1},
    \end{aligned}
\end{equation}
respectively, where $\tilde{\mathbf{R}}_k\triangleq\left(1+\gamma_k\right)\mathbf{R}_k-\gamma_k\mathbf{R}$. Resorting to the MM algorithm, a concave surrogate function $\hat{\zeta}_k\left(\mathbf{u}_{t,n}\right)$ that serves as a lower bound for \eqref{zeta} can be constructed via second-order Taylor expansion as follows:

\textit{Proposition 4:} For a given point $\mathbf{u}_{t,n}'$, a lower-bound surrogate function of $\tilde{\zeta}_k\left(\mathbf{u}_{t,n}\right)$ is given as
\begin{equation}
\begin{aligned}
    \hat{\zeta}_k\left(\mathbf{u}_{t,n}\right)&=\tilde{\zeta}_k\left(\mathbf{u}'_{t,n}\right) + \nabla \tilde{\zeta}_k\left(\mathbf{u}'_{t,n}\right)\cdot\left(\mathbf{u}_{t,n}-\mathbf{u}'_{t,n}\right) - \\
    &\enspace\enspace\frac{\delta_{t,k,n}}{2}\left\Vert\mathbf{u}_{t,n}-\mathbf{u}'_{t,n}\right\Vert_2^2 + \hat{\epsilon}_{t,k,n} \leq \tilde{\zeta}_k\left(\mathbf{u}_{t,n}\right),
\end{aligned}
\end{equation}
where the expressions of $\nabla \tilde{\zeta}\left(\mathbf{u}_{t,n}\right)$ and $\delta_{t,k,n}$ are given in Appendix \ref{APP6}, $\hat{\epsilon}_{t,k,n}$ is the constant independent of $\mathbf{u}_{t,n}$.

\textit{Proof:} See Appendix \ref{APP6}.

Overall, constraint $\textrm{C1}'''$ for given $\left\{\mathbf{u}_{t',n'},(t',n')\neq(t,n)\right\}$ is reformulated as
\begin{equation}
    \textrm{C1}'''': \hat{\zeta}_k\left(\mathbf{u}_{t,n}\right)+\nu \geq 0,\ \forall k\in\mathcal{K}.
\end{equation}

\textit{2) $\omega\left(\mathbf{u}_{t,n}\right)$ for given $\{\mathbf{u}_{t',n'},(t',n')\neq(t,n)\}$}

Based on \eqref{OBJ}, $\omega\left(\mathbf{u}_{t,n},\nu\right)$ can be reformulated as 
\begin{equation}\label{OBJ4}
\begin{aligned}
    \omega\left(\mathbf{u}_{t,n}\right)=&P_{t,0,n}\left|a_{t,0,n}\left(\mathbf{u}_{t,n}\right)\right|^2 \\
    &+ 2\mathscr{R}\left\{a^*_{t,0,n}\left(\mathbf{u}_{t,n}\right)q_{t,0,n}\right\} + \epsilon_{t,0,n}
\end{aligned}
\end{equation}
where $\epsilon_{t,0,n}$ is the term independent of $\mathbf{u}_{t,n}$, $a_{t,0,n}$ denotes the $n$-th element of $\mathbf{a}_{t,0}$, $P_{t,0,n}$ and $q_{t,0,n}$ are defined as 
\begin{equation}
    p_{t,0,n}=TN\Big(\sum_{r=1}^{M_\textrm{r}}\left[\boldsymbol{\Psi}_r\right]_{t,t}\beta_{t,r}\Big)[\mathbf{E}_t\mathbf{R}\mathbf{E}_t^H]_{n,n},
\end{equation}
and
\begin{equation}
    \begin{aligned}
        q_{t,0,n}=TN\sum_{(t',n')\neq(t,n)}&\Big(\sum_{r=1}^{M_\textrm{r}}\left[\boldsymbol{\Psi}_r\right]_{t,t'}\sqrt{\beta_{t',r}\beta_{t,r}}\Big)\\
        &[\mathbf{E}_t\mathbf{R}\mathbf{E}_{t'}^H]_{n,n'}a_{t',0,n'}\left(\mathbf{u}_{t',n'}\right),
    \end{aligned}
\end{equation}
respectively. It can be observed that the expression of $\omega\left(\mathbf{u}_{t,n}\right)$ has a similar but degenerate mathematical structure to $\tilde{\zeta}_k\left(\mathbf{u}_{t,n}\right)$, i.e., the corresponding matrix/vector degenerates to a vector/scalar. Therefore, by following a similar procedure as in \textit{Proposition 4}, we can obtain a lower-bound surrogate function of $\omega\left(\mathbf{u}_{t,n}\right)$ as
\begin{equation}
\begin{aligned}
    \hat{\omega}\left(\mathbf{u}_{t,n}\right)&=\omega\left(\mathbf{u}_{t,n}'\right) + \nabla \omega\left(\mathbf{u}_{t,n}'\right)\cdot\left(\mathbf{u}_{t,n}-\mathbf{u}_{t,n}'\right) - \\
    &\frac{\delta_{t,0,n}}{2}\left\Vert\mathbf{u}_{t,n}-\mathbf{u}_{t,n}'\right\Vert_2^2 + \hat{\epsilon}_{t,0,n} \leq \omega\left(\mathbf{u}_{t,n}\right)
\end{aligned}
\end{equation}
The specific derivations and the expressions are omitted due to space limitation.

Overall, given $\left\{\mathbf{u}_{t',n'},(t',n')\neq(t,n)\right\}$, the optimization problem with respect to $(\mathbf{u}_{t,n},\nu)$ can be reformulated as
\begin{equation}
    \left(\mathscr{P}_{3.2.t.n}\right): \mathop{\textrm{max}}\limits_{\mathbf{u}_{t,n},\nu}\ \hat{\omega}\left(\mathbf{u}_{t,n}\right)-\eta\cdot\nu\enspace\textrm{s.t.}\enspace\textrm{C1}'''',\textrm{C3},\textrm{C5},\textrm{C6},
\end{equation}
which is a quadratically constrained quadratic programming (QCQP)
problem and can be efficiently solved via CVX.

\subsection{Optimizing $(\mathbf{v},\nu)$ with given $(\mathbf{w}^{(\kappa-1)},\mathbf{s}_0^{(\kappa-1)},\mathbf{u}^{(\kappa-1)})$}\label{V.C}

Since the objective function is independent of the positions of UEs' FAs $\mathbf{v}$, the subproblem with given $(\mathbf{w}^{(\kappa-1)},\mathbf{s}_0^{(\kappa-1)},\mathbf{u}^{(\kappa-1)})$ can be formulated as
\begin{equation}
\begin{aligned}
    \left(\mathscr{P}_{3.3}\right): &\mathop{\textrm{max}}_{\mathbf{v},\nu}\ -\eta\cdot\nu\\
    \textrm{s.t.}\enspace&\textrm{C1}''':\zeta_k\left(\mathbf{v}\right)+\nu\geq 0,\ \forall k\in\mathcal{K},\textrm{C4},\textrm{C6}
\end{aligned}
\end{equation}
if the original constraint $\textrm{C1}$ is not satisfied during iterations, otherwise, the subproblem is equivalent to
\begin{equation}
    \left(\mathscr{P}'_{3.3}\right): \mathop{\textrm{find}}\ \mathbf{v}\quad
    \textrm{s.t.}\enspace\textrm{C1}'':\zeta_k\left(\mathbf{v}\right)\geq 0,\ \forall k\in\mathcal{K},\textrm{C4}
\end{equation}
based on the principle of penalty-based method. Problems $\left(\mathscr{P}_{3.3}\right)$ and $\left(\mathscr{P}'_{3.3}\right)$ can both be viewed as
a generalized feasibility problem that can only produce a solution (if the original problem is feasible) satisfying constraints $\textrm{C1}$ and $\textrm{C4}$. However, according to the impact of UEs' FA positions on the sensing performance analyzed in \textit{Remark 2}, we propose to maximize the SINR of UEs by optimizing $\mathbf{v}$, instead of merely finding a feasible solution as in previous works such as~\cite{10696953}. Furthermore, it can be observed that the position of UE $k$'s FA has no effect on other UEs' performance. Therefore, the optimization of the positions of different UEs' FAs can be implemented in parallel and the subproblem corresponding to UE $k$ can be formulated as
\begin{equation}
    \left(\mathscr{P}_{3.3.k}\right): \mathop{\textrm{max}}_{\mathbf{v}_k}\ \textrm{SINR}_k\left(\mathbf{v}_k\right)\quad
    \textrm{s.t.}\enspace\textrm{C4}':\mathbf{v}_k\in\mathcal{C}^\textrm{r}.
\end{equation}

Although the method proposed in Section \ref{V.B} can be similarly employed to transform the objective function of problem  $\left(\mathscr{P}_{3.3.k}\right)$ into a concave one and solve it iteratively via the CVX toolbox, an efficient GA-based scheme with much lower complexity is proposed to find a sub-optimal solution. 

According to \eqref{h_tk} and \eqref{SINR}, and defining $N_k(\mathbf{v}_k)\triangleq\mathbf{b}^H_{k}\left(\mathbf{v}_k\right)$ $\bar{\mathbf{R}}_k\mathbf{b}_{k}\left(\mathbf{v}_k\right)$, $D_k(\mathbf{v}_k)\triangleq\mathbf{b}^H_{k}\left(\mathbf{v}_k\right)(\bar{\mathbf{R}}-\bar{\mathbf{R}}_k)\mathbf{b}_{k}\left(\mathbf{v}_k\right) + \sigma_k^2$, $\textrm{SINR}_k\left(\mathbf{v}_k\right)$ can be reformulated as
\begin{equation}
\textrm{SINR}_k\left(\mathbf{v}_k\right)=N_k(\mathbf{v}_k)/D_k(\mathbf{v}_k),
\end{equation}
where $\mathbf{b}_{k}\left(\mathbf{v}_k\right) = [\mathbf{b}^H_{1,k}\left(\mathbf{v}_k\right),\dots,\mathbf{b}^H_{M_\textrm{t},k}\left(\mathbf{v}_k\right)]^H$, $\bar{\mathbf{R}}_k=\boldsymbol{\Sigma}_{k}\mathbf{A}_k$\\$ \mathbf{R}_k\mathbf{A}_k^H\boldsymbol{\Sigma}_{k}^H$, $\bar{\mathbf{R}}=\boldsymbol{\Sigma}_{k}\mathbf{A}_k\mathbf{R}\mathbf{A}_k^H\boldsymbol{\Sigma}_{k}^H$, $\boldsymbol{\Sigma}_{k} = \textrm{blkdiag}\left(\boldsymbol{\Sigma}_{1,k},\dots,\right.$ $\left.\boldsymbol{\Sigma}_{M_\textrm{t},k}\right)$, $\mathbf{A}_{k} = \textrm{blkdiag}\left(\mathbf{A}_{1,k},\dots,\mathbf{A}_{M_\textrm{t},k}\right)$. Given that $\nabla_{\mathbf{v}_k} b_{t,k,l}\left(\mathbf{v}_k\right)= -j2\pi/\lambda b_{t,k,l}\left(\mathbf{v}_k\right)\boldsymbol{\varpi}_{t,k,l}$, where $b_{t,k,l}\left(\mathbf{v}_k\right)$ is the $l$-th element of $\mathbf{b}_{t,k}$, $\partial\mathbf{b}_k\left(\mathbf{v}_k\right)/\partial\mathbf{v}_k$ is expressed as
\begin{equation}
    \frac{\partial\mathbf{b}_k\left(\mathbf{v}_k\right)}{\partial \mathbf{v}_k}=-j\frac{2\pi}{\lambda}\begin{bmatrix}
b_{1,k,1}\left(\mathbf{v}_k\right)\boldsymbol{\varpi}_{1,k,1} \\
\vdots \\
b_{M_\textrm{t},k,L}\left(\mathbf{v}_k\right)\boldsymbol{\varpi}_{M_\textrm{t},k,L}
\end{bmatrix}\in\mathbb{C}^{M_\textrm{t}L\times 2}.
\end{equation}
Then, the gradient  $\nabla_{\mathbf{v}_k}\textrm{SINR}\left(\mathbf{v}_k\right)$ can be expressed as
\begin{equation}
\begin{aligned}
   &\nabla_{\mathbf{v}_k}\textrm{SINR}_k\left(\mathbf{v}_k\right) =\\
   &\qquad\frac{\nabla_{\mathbf{v}_k} N_k(\mathbf{v}_k)\cdot D_k(\mathbf{v}_k)-\nabla_{\mathbf{v}_k} D_k(\mathbf{v}_k)\cdot N_k(\mathbf{v}_k)}{D_k^2(\mathbf{v}_k)},
\end{aligned}
\end{equation}
where 
\begin{equation}
    \nabla_{\mathbf{v}_k} N_k(\mathbf{v}_k)=2\mathscr{R}\Big\{\big(\frac{\partial\mathbf{b}_k\left(\mathbf{v}_k\right)}{\partial \mathbf{v}_k}\big)^H\bar{\mathbf{R}}_k\mathbf{b}_k\left(\mathbf{v}_k\right)\Big\},
\end{equation}

\begin{equation}
    \nabla_{\mathbf{v}_k} D_k(\mathbf{v}_k)=2\mathscr{R}\Big\{\big(\frac{\partial\mathbf{b}_k\left(\mathbf{v}_k\right)}{\partial \mathbf{v}_k}\big)^H\big(\bar{\mathbf{R}}-\bar{\mathbf{R}}_k\big)\mathbf{b}_k\left(\mathbf{v}_k\right)\Big\}.
\end{equation}

Overall, according to GA algorithm, $\mathbf{v}_k$ can be iteratively updated as
\begin{equation}
    \mathbf{v}_k^{(\varsigma)}=\mathbf{v}_k^{(\varsigma-1)} + \theta^{(\varsigma-1)}\nabla_{\mathbf{v}_k}\textrm{SINR}_k\big(\mathbf{v}_k^{(\varsigma-1)}\big),
\end{equation}
where $(\varsigma)$ denotes the iteration time of GA algorithm, $\theta^{(\varsigma)}$ denotes the step size in each iteration. The most widely-used step size policy for non-convex problem is decreasing the step size as iteration proceeds. Under the premise of satisfying the constraints $\textrm{C4}'$, we adopt $\theta^{(\varsigma)}=a/(\varsigma+b)$ as in~\cite{10596930}, where $a$ and $b$ are predefined constants for convergence.

\begin{figure}[htbp] 
    \vspace{-10pt}
    \centering 
    \includegraphics[width=0.38\textwidth]{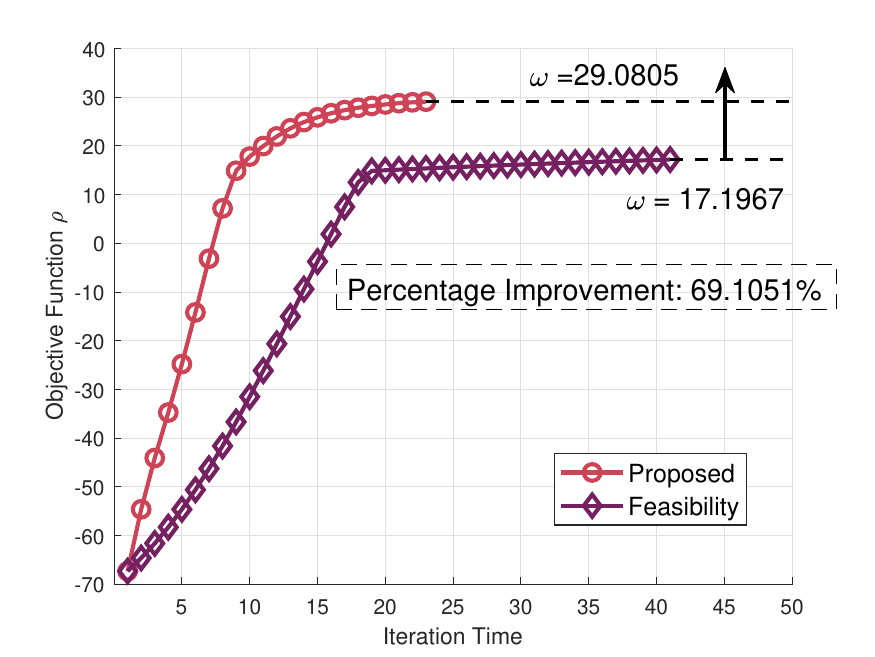} 
    \caption{Objective function versus iteration time under different UEs' FA positions optimization schemes.} 
    \label{Fig2} 
    \vspace{-10pt}
\end{figure}

Fig. \ref{Fig2} demonstrates the superiority of the proposed GA-based scheme over the benchmark scheme of merely solving the feasibility problem, based on the system configuration detailed in Section \ref{NLS}. In this setup, only the transmit beamforming and UEs' FA positions are optimized, while the ISAC transmitters' FAs remain fixed. It can be observed from both the curve slopes and the required iteration time that the proposed scheme can converge significantly faster. Furthermore, it can achieve an approximate 69.1\% improvement over the benchmark scheme, which is attributed to the full exploitation of the mechanism discussed in \textit{Remark 2}.

\subsection{Overall Algorithm, Convergence Analysis, and Computational Complexity}\label{A,C,C}

The detailed procedures of the proposed overall solution are summarized in Algorithm \ref{alg}. Benefiting from the proposed penalty-based framework, the initial positions $\{\mathbf{u}^{(0)}_{t,n}\}$ can be straightforwardly initialized using the simple CP method proposed in \cite{10243545}, while $\{\mathbf{v}^{(0)}_k\}$ is set to the center of the considered region $\mathcal{C}^\textrm{r}$. Crucially, this initialization does not strictly require constraint C1 to be satisfied.

\vspace{-3pt}
\begin{algorithm}[htbp]\small
    \caption{Overall Algorithm for Solving Problem $(\mathscr{P}_1)$}\label{alg}
    
    \SetKwInput{Initialize}{Initialize}
    
    \Initialize{$\{\mathbf{u}^{(0)}_{t,n}\}$, $\{\mathbf{v}^{(0)}_k\}$, $\eta$, $\theta^{(0)}$, $\epsilon$, $\kappa = 0$.}
    
    \Repeat{the fractional increase of $\rho^{(\kappa)}=\omega^{(\kappa)} - \eta\cdot\nu^{(\kappa)}$ is below the predefined threshold $\epsilon$}{
        $\kappa = \kappa + 1$.
        
        Update $\{\mathbf{R}^{(\kappa)}_k\}, \mathbf{R}^{(\kappa)}_0$ by solving problem $(\mathscr{P}_{3.1}')$.
        
        Update $\{\mathbf{u}^{(\kappa)}_{t,n}\}$ by sequentially solving problem $(\mathscr{P}_{3.2.t.n})$.
        
        Update $\{\mathbf{v}^{(\kappa)}_{k}\}$ by solving problem $(\mathscr{P}_{3.3.k})$ in parallel.
    }
    
    \If{$\nu^{(\kappa)}$ converges to a near-zero value}{
        The final transmit beamforming $\{\mathbf{w}^*_k\}$ and dedicated sensing signal $\mathbf{s}^*_0$ can be obtained through SVD of $\{\mathbf{R}^{(\kappa)}_k\}$ and $\mathbf{R}^{(\kappa)}_0$, respectively.
    }
    \Else{
        The original problem is infeasible.
    }
\end{algorithm}
\vspace{-7pt}

\textit{Convergence Analysis:} During the AO procedure in Algorithm \ref{alg}, each subproblem is solved either optimally or to a local optimum, ensuring that the value of objective function is monotonically non-decreasing. Given that the system utility is upper-bounded by the finite available resources (e.g., transmit power and spatial DoFs), the proposed Algorithm \ref{alg} is guaranteed to converge.

\textit{Computational Complexity:} The interior-point method is employed to solve the formulated subproblems. Specifically, in step 3, problem $\left(\mathscr{P}_{3.1}'\right)$ is a semidefinite program (SDP). Assuming all matrix variables are dense without special structures, the worst-case computational complexity of solving problem $\left(\mathscr{P}_{3.1}'\right)$ is given by $\mathcal{O}\big(\left(NM_\textrm{t}\right)^{4.5}\textrm{log}\left(1/\epsilon_1\right)\big)$, where $\epsilon_1\geq0$ denotes the solution accuracy~\cite{5447068}. In step 4, the QCQP $\left(\mathscr{P}_{3.2.t.n}\right)$ can be first reformulated as a standard second-order cone program (SOCP) featuring $K+1$ SOC constraints of dimension 4. Such a problem can be efficiently solved with computational complexity $\mathcal{O}\big(4^3\cdot\left(K+1\right)\big)$~\cite{lobo1998applications}. Consequently, the total computational complexity of sequentially solving $\left(\mathscr{P}_{3.2.t.n}\right)$ is given by $\mathcal{O}\left(64\cdot\mathcal{I}_1NM_\textrm{t}\left(K+1\right)\right)$, where $\mathcal{I}_1$ denotes the number of iterations required for convergence in step 4. In step 5, the computational complexity is dominated by the gradient calculation, which scales as $\mathcal{O}\big(\left(LM_\textrm{t}\right)^2\big)$. Thus, the total computational complexity of solving $\left(\mathscr{P}_{3.3.k}\right)$ is $\mathcal{O}\big(\mathcal{I}_2K\left(LM_\textrm{t}\right)^2\big)$, where $\mathcal{I}_2$ denotes the number of iterations required for convergence in step 5. Finally, the overall computational complexity of Algorithm \ref{alg} is represented by $\mathcal{O}\big(\mathcal{I}\big(\left(NM_\textrm{t}\right)^{4.5}\textrm{log}\left(1/\epsilon_1\right)+64\cdot\mathcal{I}_1NM_\textrm{t}\left(K+1\right)\big.\big.$ $\big.\big.+\mathcal{I}_2K\left(LM_\textrm{t}\right)^2\big)\big)$, where $\mathcal{I}$ denotes the number of iterations required for the AO procedure.

\section{Numerical Results}
In this section, numerical results are presented to validate the effectiveness of the proposed scheme. First, the general setup of simulation parameters is presented in Section \ref{SP}. Then, the convergence behavior is illustrated in Section \ref{CB}. Afterwards, motivated by the discussion in \textit{Remark 2}, we consider two classic resource-limited scenarios-namely, noise-limited and interference-limited scenarios-in Sections \ref{NLS} and \ref{ILS}, respectively, to demonstrate the superiority of the DS-FAS-assisted multi-static ISAC architecture. Finally, the impacts of moving region sizes on both sides are investigated in Section \ref{IoMRaNoP}.

\subsection{Simulation Parameters}\label{SP}
Unless otherwise specified, the simulation parameters are set as follows. A circular region with radius of $200$ m is considered, where $M_\textrm{t}+M_\textrm{r}=6$ APs and $K=4$ UEs are uniformly distributed. Following the approach in~\cite{10494224}, the $M_\textrm{r}=2$ APs situated closest to the target are designated as sensing receivers, while the remaining $M_\textrm{t}=4$ APs serve as ISAC transmitters. Each AP is equipped with $N=4$ FAs, while each UE is equipped with a single FA. Without loss of generality, we assume that all communication channels have an identical number of paths, i.e., $L=12$. The moving regions for each transmit FA and each UE's FA are constrained within $X^\textrm{t}=Y^\textrm{t}=2\lambda$ and  $X^\textrm{r}=Y^\textrm{r}=\lambda$, respectively. The elements of PRM are modeled as $\upsilon_{t,k,l}\sim\mathcal{CN}\big(0,K_0d_{t,k}^{-\iota}/L\big)$, where $K_0=-40\textrm{dB}$ denotes the path loss at the reference distance of 1 m, $d_{t,k}$ represents the distance between ISAC transmitter $t$ and UE $k$, and $\iota=2.8$~\cite{11033708}. The path loss of sensing channel is modeled as $\lambda^2/\big(\left(4\pi\right)^3d_{t,0}^2d_{r,0}^2\big)$~\cite{richards2010principles}, where $d_{t,0}$ and $d_{r,0}$ denote the distance from ISAC transmitter $t$ to target and from the target to sensing receiver $r$, respectively. The RCS are assumed to be independent, with the variance set as $\sigma^2_{t,r}=1$. The noise variance at S\&C receivers are set as  $\sigma^2_k=\sigma_r^2=-95\textrm{dBm}$. The power budget per ISAC transmitter is set as $P_t=20\textrm{dBm}$. The convergence threshold of Algorithm \ref{alg} is set as $\epsilon=1e^{-2}$.

\subsection{Convergence Behavior}\label{CB}
\begin{figure*}[htbp]
\vspace{0pt}
\centering
\subfloat[Algorithm \ref{alg}]{\includegraphics[width=0.28\linewidth]{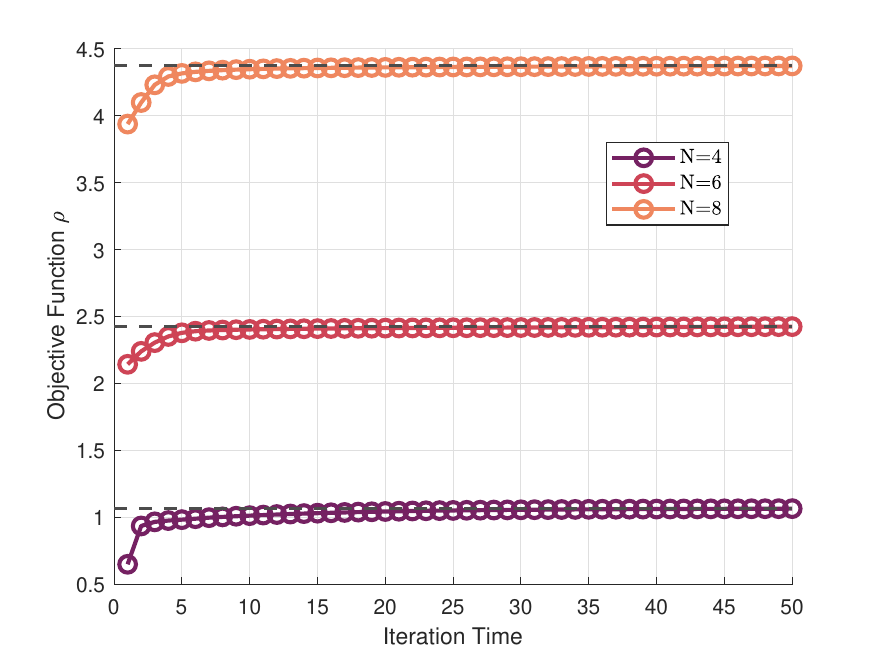}%
\label{Fig3.a}}
\subfloat[Subproblem $\left(\mathscr{P}_{3.2}\right)$]
{\includegraphics[width=0.28\linewidth]{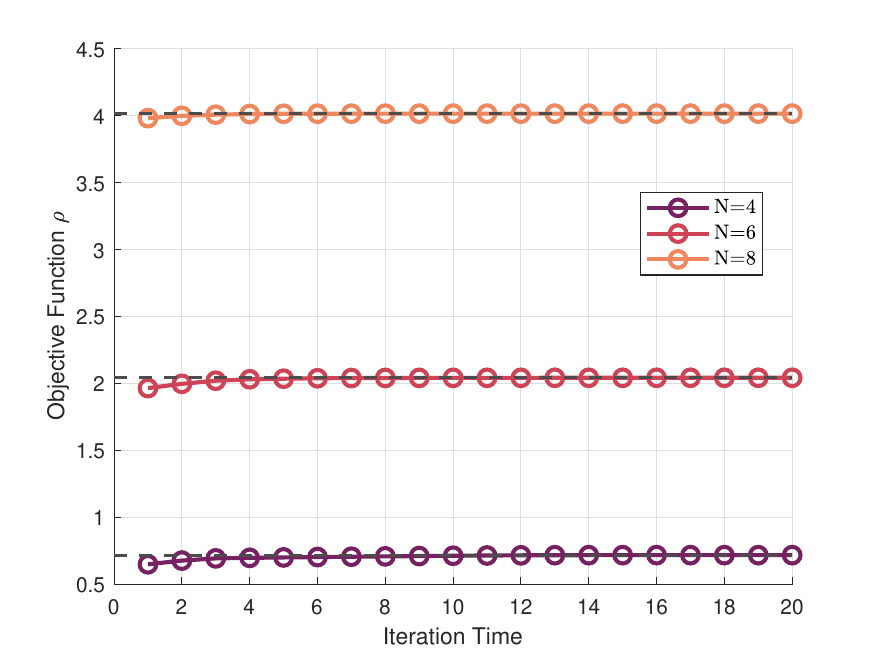}%
\label{Fig3.b}}
\subfloat[Subproblem $\left(\mathscr{P}_{3.3.k}\right)$]
{\includegraphics[width=0.28\linewidth]{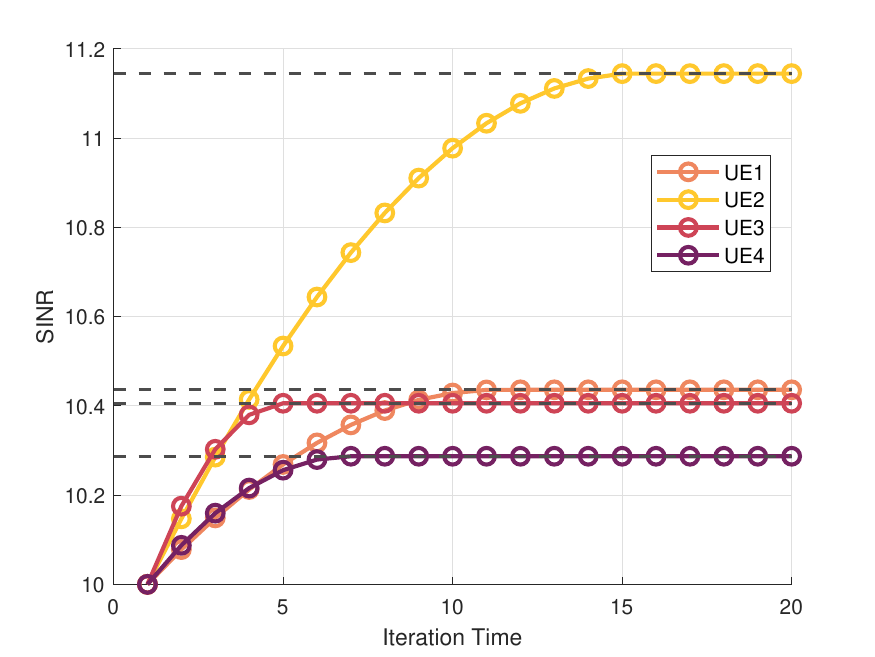}%
\label{Fig3.c}}
\caption{The convergence behavior of the overall algorithm and solving different subproblems.}
\vspace{-10pt}
\label{Fig3}
\end{figure*}

Fig. \ref{Fig3} illustrates the convergence behavior of Algorithm \ref{alg}, along with that of solving subproblems $\left(\mathscr{P}_{3.2}\right)$ and $\left(\mathscr{P}_{3.3.k}\right)$, for various numbers of FAs at the APs, i.e., $N\in\left\{4,6,8\right\}$. In Fig. \ref{Fig3.a}, it can be observed that the objective function increases monotonically and Algorithm \ref{alg} converges within approximately 20-30 iterations, irrespective of the number of FAs $N$ at the APs. In Fig. \ref{Fig3.b}, it can be observed that the procedure for solving subproblem $\left(\mathscr{P}_{3.2}\right)$ can converge within approximately 5-10 iterations for different $N$. The SINR values of the four UEs versus the number of iterations are illustrated in Fig. \ref{Fig3.c}. It can be observed that the GA-based procedure for solving subproblem $\left(\mathscr{P}_{3.3.k}\right)$ converges within approximately 15-20 iterations. These simulation results are consistent with the convergence analysis in Section \ref{A,C,C}.

\subsection{Noise-Limited Scenario}\label{NLS}
In this subsection, the noise variances are increased to $\sigma^2_k=\sigma_r^2=-85\textrm{dBm}$ to simulate the noise-limited scenario. Under this setup, more transmit power is required to satisfy the stringent SINR requirement $\gamma_k$, thereby intensifying the resource competition and conflict between the S\&C functions. For comparison, the following existing schemes are considered as baselines in the remainder of this section:
\begin{itemize}
\item \textbf{Transmit FA only (T-FAS):} Only the ISAC transmitters are equipped with FAs, while each UE's antenna is fixed at the center of its respective region.

\item \textbf{Receive FA only (R-FAS):} Only the UEs are equipped with FAs, while the antennas at each ISAC transmitter are fixed at positions determined by the CP method.

\item \textbf{FPA-ULA:} All antenna positions are fixed. Specifically, each UE's antenna is located at the center of the their respective region, while the antennas at each ISAC transmitter are arranged as ULAs.

\item \textbf{FPA-CP:} All antenna positions are fixed. Specifically, each UE's antenna is located at the center of the their respective region, while the antenna positions at each ISAC transmitter are determined by the CP method.
\end{itemize}

\begin{figure}[htbp] 
\vspace{-5pt}
    \centering 
    \includegraphics[width=0.38\textwidth]{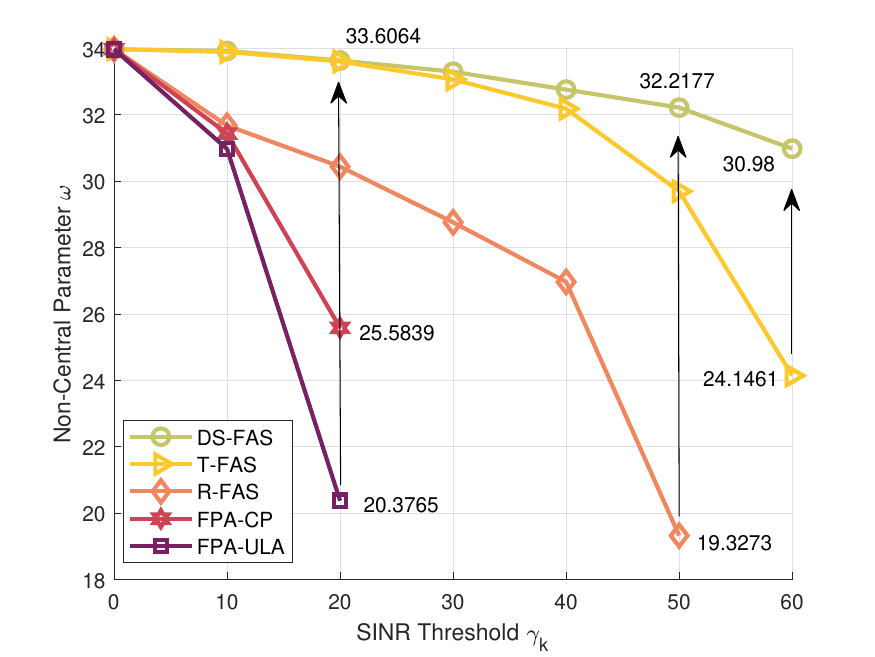} 
    \caption{Non-central parameter $\omega$ versus SINR threshold $\gamma_k$ in a specific noise-limited scenario.} 
    \label{Fig4} 
\vspace{-15pt}
\end{figure}

Considering a random realization of system topology, Fig. \ref{Fig4} compares the proposed scheme (denoted as “\textbf{DS-FAS}”) with aforementioned baseline schemes under progressively stringent SINR requirements. When $\gamma_k=0$, the sensing-only scenario is considered, and all schemes exhibit identical sensing performance, which aligns with the theoretical analysis presented in \textit{Proposition 3}. However, as $\gamma_k$ increases, the advantage of FAS-assisted system becomes evident. Specifically, while the FPA-based system fails to satisfy the SINR requirement at $\gamma_k=30$, the FAS-assisted system can accommodate SINR constraints of at least $\gamma_k=50$, even under the SS-FAS configuration. Furthermore, the performance gap between the FAS-assisted system and FPA-based system becomes increasingly prominent as the SINR requirement tightens. Notably, at $\gamma_k=20$, the proposed scheme achieves approximately 31.36\% and 64.93\% gains in $\omega$ compared to the “FPA-CP” and “FPA-ULA” baselines, respectively. Regarding the SS-FAS-assisted system, the proposed scheme maintains a substantial performance margin compared to the “R-FA” baseline, achieving a 66.70\% gain in $\omega$ at $\gamma_k=50$. In contrast, the “T-FA” baseline exhibits a smaller performance gap relative to the proposed scheme, particularly under lower SINR requirements, and it significantly outperforms the “R-FA” baseline as $\gamma_k$ increases. This is because the multiple FAs at each ISAC transmitter exploit enhanced spatial DoFs compared to the “R-FA” baseline in the considered scenario. Nevertheless, by leveraging DS-FAS, the proposed scheme still maintains a 28.3\% gain in $\omega$ at $\gamma_k=60$ compared to the “T-FA” baseline. It is also worth noting that the “T-FA” baseline cannot consistently outperform the “R-FA” baseline across all regimes, which will be further elaborated in Section~\ref{ILS}.

\begin{figure}[htbp] 
    \vspace{-15pt}
    \centering 
    \includegraphics[width=0.38\textwidth]{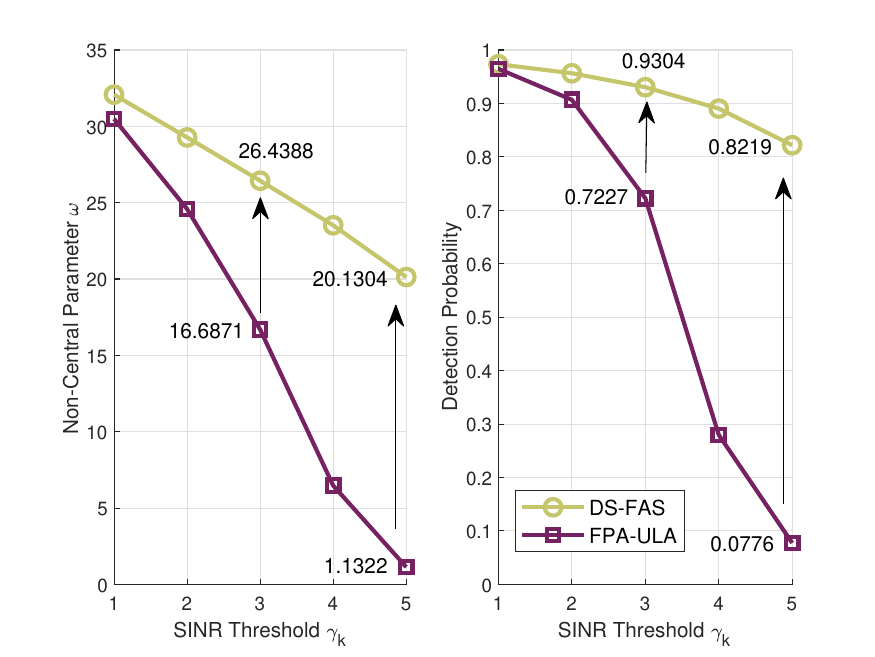} 
    \caption{Average non-central parameter $\omega$ and detection probability versus SINR threshold $\gamma_k$ in the generalized noise-limited scenario.} 
    \label{Fig5} 
    \vspace{-15pt}
\end{figure}

To demonstrate the robustness of the proposed scheme with respect to the system topology, the non-central parameter $\omega$ is computed over 200 random realizations of the system topology, and the corresponding detection probability is also evaluated with the false alarm probability fixed at 5\%. However, since distinct system topologies exhibit significant variations in path loss due to the large radius of the considered area, it is challenging to set a universal SINR threshold for different Monte Carlo trials that consistently represents the noise-limited scenarios. To address this issue, we normalize the corresponding parameters to a specific interval as in \cite{10839251}. Specifically, the elements of PRM are modeled as $\delta_{t,k,l}\sim\mathcal{CN}\left(0,1/L_{t,k}\right)$, while the path loss of sensing channel is set as 0.05 without loss of generality. Furthermore, the average SNR is assumed to be $\textrm{SNR}=P_\textrm{t}/\sigma_{k/ r}^2=4$, and the SINR constraint $\gamma_k$ varies within the set $\{1,2,3,4,5\}$. In cases where the SINR constraints are infeasible, the non-centrality parameter is set to zero to represent a failure of the ISAC service. A similar performance superiority to that in Fig.~\ref{Fig4} can be observed from Fig.~\ref{Fig5}. Specifically, the proposed scheme achieves approximately 58.44\% and 1677.99\% average gains in $\omega$ at $\gamma_k=3$ and $\gamma_k=5$, respectively, compared with the “FPA-ULA” baseline. Correspondingly, improvements of up to 20.77\% and 74.43\% in detection probability can be achieved. 

\subsection{Interference-Limited Scenario}\label{ILS}
\begin{figure}[htbp] 
\vspace{-5pt}
    \centering 
    \includegraphics[width=0.38\textwidth]{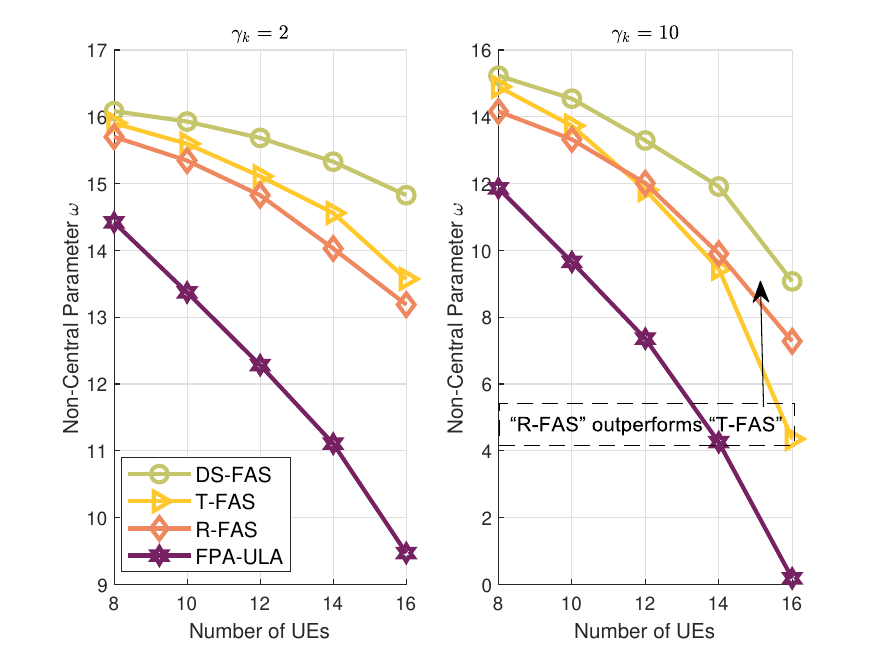} 
    \caption{Average non-central parameter $\omega$ versus the number of UEs $K$ in the generalized interference-limited scenario under varied SINR thresholds $\gamma_k$.} 
    \label{Fig6} 
    \vspace{-5pt}
\end{figure}

In this subsection, considering a total of $NM_\textrm{t}=16$ transmit FAs are deployed, we gradually increase the number of UEs from 8 to 16, i.e., $K\in\{8,10,12,14,16\}$, to simulate increasingly interference-limited scenarios with SINR thresholds $\gamma\in\{2,10\}$, and the non-central parameter $\omega$ is averaged over 200 Monte Carlo trials. It can be observed from Fig.~\ref{Fig6} that the proposed scheme can achieve significant performance gains compared to the “FPA-ULA” baseline and consistently outperforms the “T-FA” and “R-FA” baselines, leveraging the additional DoFs afforded by the DS-FAS. More importantly, the performance gap between the proposed scheme and the baselines widens as the number of UEs increases. Unlike the noise-limited scenario, the DoFs provided by multiple antennas become increasingly scarce as the number of UEs grows, causing the interference term to become dominant in the SINR expression. Consequently, even with a relatively low SINR threshold, the aforementioned performance gap remains prominent, as illustrated in the left subfigure of Fig.~\ref{Fig6}. Another interesting observation from the right subfigure of Fig.~\ref{Fig6} is that the “R-FA” baseline gradually surpasses the “T-FA” baseline as the interference becomes more severe, which differs from the trend observed in Fig.~\ref{Fig4} in Section \ref{NLS}. This occurs because the transmit FAs are shared resources that must simultaneously satisfy the SINR requirements of all UEs and enhance the sensing task. As the number of UEs increases, it becomes increasingly difficult to find a transmit positions that better serve all entities. In contrast, each receive FA is dedicated to a single UE, thus, it's spatial flexibility remains unconstrained by the growing number of UEs.

\textit{Lessons Learned:} Based on the simulation results and analyses in Sections \ref{NLS} and \ref{ILS}, several practical insights can be drawn. First, in the interference-limited scenario, deploying FAs exclusively at the receiver side may be preferred. As demonstrated in Fig.~\ref{Fig6}, this configuration can achieve performance comparable to that of the DS-FAS assisted system while bypassing the high computational overhead associated with optimizing transmit FA positions. On the contrary, in the noise-limited scenario with an abundant number of antennas, deploying FAs at the transmitter side is preferred despite the higher computational complexity. As demonstrated in Fig.~\ref{Fig4}, this configuration can achieve significant performance gains over both the FPA-assisted system and system equipped with FAs only at the receiver side. Furthermore, under more stringent SINR requirements in the noise-limited scenario, additionally deploying FAs at the receiver side is highly recommended. The proposed GA-based optimization for receive FAs not only achieves further performance enhancements unattainable by the 'T-FA' baseline alone as illustrated in Fig.~\ref{Fig4} with $\gamma_k=60$, but also entails a remarkably low computational complexity.

\subsection{Impact of Moving Region Sizes on Both Sides}\label{IoMRaNoP}
\begin{figure}[htbp] 
\vspace{-5pt}
    \centering 
    \includegraphics[width=0.35\textwidth]{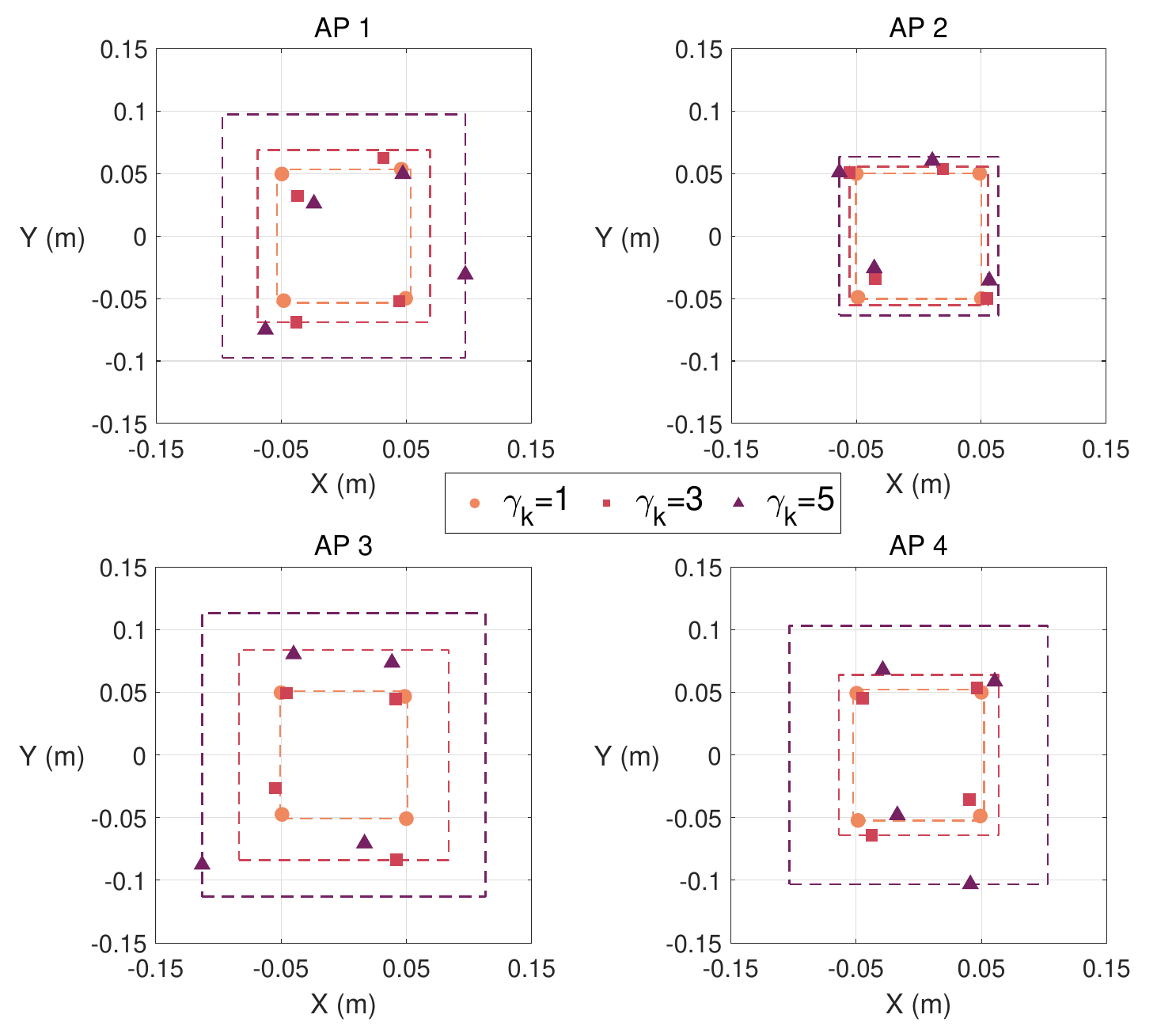} 
    \caption{Transmit FA positions under varied SINR thresholds.} 
    \label{Fig7} 
\vspace{-20pt}
\end{figure}

\begin{figure}[htbp] 
\vspace{0pt}
    \centering 
    \includegraphics[width=0.35\textwidth]{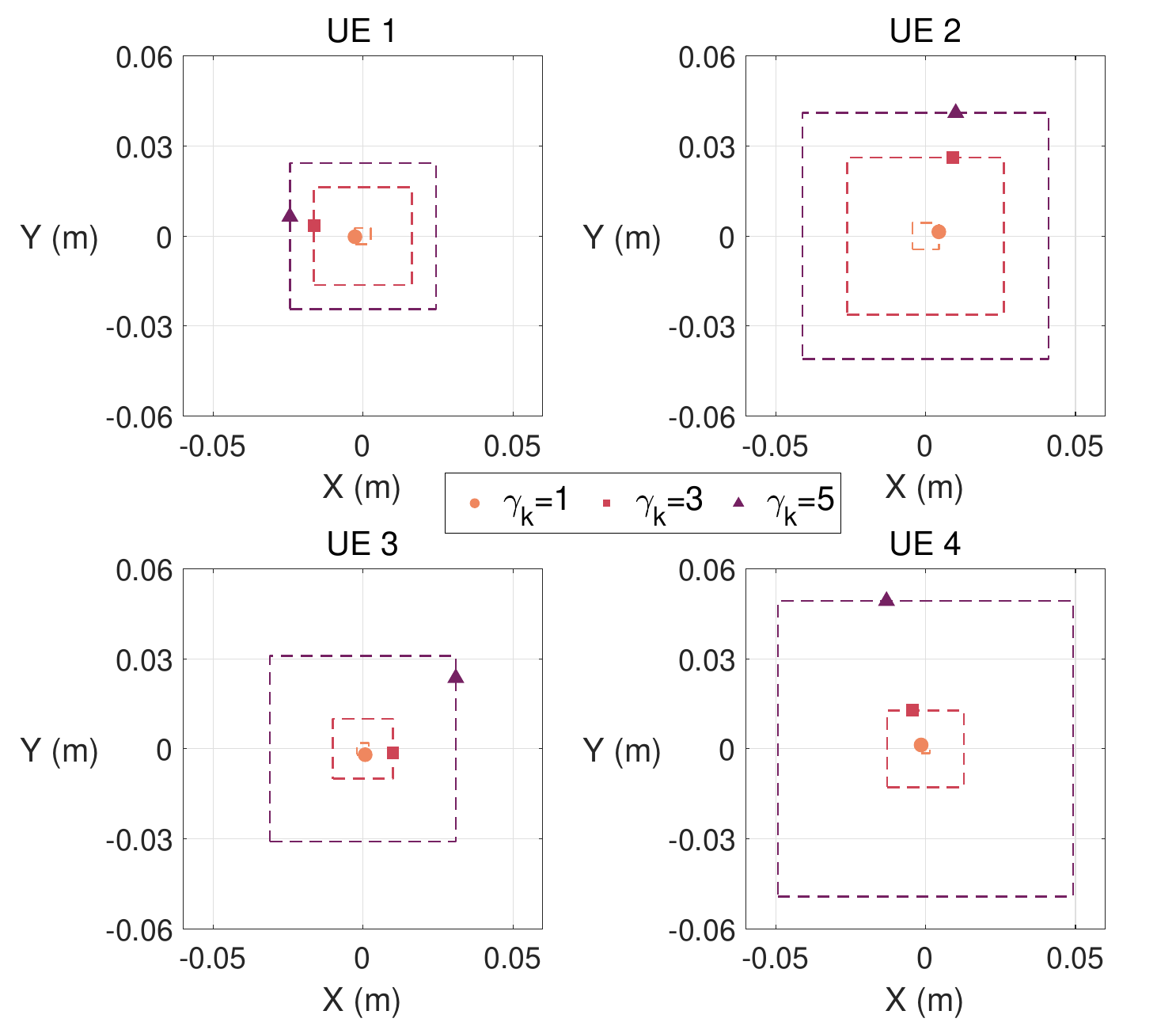} 
    \caption{Receive FA positions under varied SINR thresholds.} 
    \label{Fig8} 
\vspace{-10pt}
\end{figure}

Finally, the impact of the moving region sizes on both sides on the system performance is investigated in this subsection. A system setup similar to that in Section \ref{NLS} is employed with a random channel realization and the SINR constraint $\gamma_k$ varies within the set $\{1,3,5\}$. To provide a more intuitive illustration, the optimized FA positions at both the ISAC transmitters and the UEs are visualized in Fig.~\ref{Fig7} and~\ref{Fig8}, respectively, where the boundaries of the needed corresponding moving regions are denoted by dotted lines. To ensure a fair comparison across different $\gamma_k$, the FAs at the ISAC transmitters are initialized at the vertices of a square with a side length of $0.05$ m to satisfy the minimum distance constraint (C5) initially. Meanwhile, the FAs at the UEs are initialized at the origin of their respective moving regions. It can be observed that with a more stringent SINR constraint, larger moving regions on both sides are required to offer more DoFs for exploring better channel conditions. 

\section{Conclusion}
This paper has investigated a novel DS-FAS-assisted multi-static ISAC system, where FA positions and transmit beamforming are jointly optimized to maximize target detection probability while adhering to the communication QoS requirements. A penalty-based approach is utilized to reformulate the original problem into a robust framework that circumvents initial feasibility. Then, an AO-based algorithm is developed to iteratively optimize the decoupled variables. Notably, the receivers' FA positioning subproblem is transformed from the feasibility problem to a SINR maximization task, which significantly reduces computational complexity while fully exploiting the spatial DoFs at the receiver side. Simulation results validate the superiority of the proposed DF-FAS-assisted multi-static ISAC system. 

\begin{appendices}
   \setcounter{equation}{0}
    \renewcommand\theequation{A.\arabic{equation}}
    \section{Proof of Proposition 1} \label{APP1}
    Given $\boldsymbol{\alpha}$ and that hypothesis $\mathcal{H}_1$ holds, the conditional probability density function (C-PDF) of $\mathbf{y}$ is given as
    \begin{equation}
        f\left(\mathbf{y};\boldsymbol{\alpha},\mathcal{H}_1\right) = \frac{1}{(\pi\sigma^2)^{NM_\textrm{r}T}}\exp\Big(-\frac{1}{\sigma^2}\big\Vert\mathbf{y}-{\tilde{\mathbf{G}}}\boldsymbol{\alpha}\big\Vert^2\Big).
    \end{equation}
    Similarly, given that hypothesis $\mathcal{H}_0$ holds, the C-PDF of $\mathbf{y}$ is given as
    \begin{equation}
        f\left(\mathbf{y};\mathcal{H}_0\right) = \frac{1}{(\pi\sigma^2)^{NM_\textrm{r}T}}\exp\Big(-\frac{1}{\sigma^2}\left\Vert\mathbf{y}\right\Vert^2\Big).
    \end{equation}

    According to \cite{kay1998fundamentals}, the GLR is given as $L_\textrm{G}(\mathbf{y})=f\left(\mathbf{y};\hat{\boldsymbol{\alpha}},\mathcal{H}_1\right)/{f\left(\mathbf{y};\mathcal{H}_0\right)}$, where $\hat{\boldsymbol{\alpha}}$ is the maximum likelihood estimation (MLE) of $\boldsymbol{\alpha}$, calculated as $\hat{\boldsymbol{\alpha}} = (\tilde{\mathbf{G}}^H\tilde{\mathbf{G}})^{-1}\tilde{\mathbf{G}}^H\mathbf{y}$.
    Accordingly, the expression of GLR can be reformulated as 
    \begin{equation}\label{LG}
    \begin{aligned}
        L_\textrm{G}(\mathbf{y}) &=\exp\big(\frac{1}{\sigma^2}\mathbf{y}^H\tilde{\mathbf{G}}(\tilde{\mathbf{G}}^H\tilde{\mathbf{G}})^{-1}\tilde{\mathbf{G}}^H\mathbf{y}\big)
    \end{aligned}
    \end{equation}

   \setcounter{equation}{0}
    \renewcommand\theequation{B.\arabic{equation}}
    \section{Proof of Proposition 2} \label{APP2}
    According to Appendix \ref{APP1}, given $\boldsymbol{\alpha}$ and that hypothesis $\mathcal{H}_1$ holds, $\hat{\boldsymbol{\alpha}}\sim\mathcal{CN}\big(\boldsymbol{\alpha},\sigma^2(\tilde{\mathbf{G}}^H\tilde{\mathbf{G}})^{-1}\big)$. Then, according to \cite{kay1998fundamentals} and \eqref{LG}, we have
    \begin{equation}
    \textrm{ln}\left(L_\textrm{G}(\mathbf{y})\right)=\frac{1}{\sigma^2}\hat{\boldsymbol{\alpha}}^H\tilde{\mathbf{G}}^H\tilde{\mathbf{G}}\hat{\boldsymbol{\alpha}}\sim\tilde{\chi}_{2M_\textrm{r}M_\textrm{t}}^2\left(\omega\right),
    \end{equation}
    where $\omega = \boldsymbol{\alpha}^H\tilde{\mathbf{G}}^H\tilde{\mathbf{G}}\boldsymbol{\alpha}/{\sigma^2}$.

    Then, given  that hypothesis $\mathcal{H}_0$ holds, $\hat{\boldsymbol{\alpha}}\sim\mathcal{CN}\big(\boldsymbol{0},\sigma^2\big.$ $\big.(\tilde{\mathbf{G}}^H\tilde{\mathbf{G}})^{-1}\big)$. Similarly, according to \cite{kay1998fundamentals} and \eqref{LG}, we have
    \begin{equation}
    \textrm{ln}\left(L_\textrm{G}(\mathbf{y})\right)=\frac{1}{\sigma^2}\hat{\boldsymbol{\alpha}}^H\tilde{\mathbf{G}}^H\tilde{\mathbf{G}}\hat{\boldsymbol{\alpha}}\sim\chi_{2M_\textrm{r}M_\textrm{t}}^2.
    \end{equation}

   \setcounter{equation}{0}
    \renewcommand\theequation{C.\arabic{equation}}
    \section{Reformulation of $\omega$}\label{APP3}
    Given the expression of $\omega$, we first reformulated $\tilde{\mathbf{G}}^H\tilde{\mathbf{G}}$ as
    \begin{equation}
    \begin{aligned}
    &\tilde{\mathbf{G}}^H\tilde{\mathbf{G}}=\sum\nolimits_{\tau=1}^T\left(\mathbf{I}_{M_\textrm{r}}\otimes\mathbf{X}[\tau]\right)^H\mathbf{G}^H\mathbf{G}\left(\mathbf{I}_{M_\textrm{r}}\otimes\mathbf{X}[\tau]\right) \\
    &=\sum_{\tau=1}^T\textrm{blkdiag}\left(\mathbf{X}[\tau]^H\mathbf{G}_1^H\mathbf{G}_1\mathbf{X}[\tau],\cdots,\mathbf{X}[\tau]^H\mathbf{G}_{M_\textrm{r}}^H\mathbf{G}_{M_\textrm{r}}\mathbf{X}[\tau]\right)
    \end{aligned}
    \end{equation}
    Based on the block-diagonal structure of $\tilde{\mathbf{G}}^H\tilde{\mathbf{G}}$, $\omega$ can be reformulated as
    \begin{equation}\label{OBJ2}
    \begin{aligned}
    \omega&=\sum_{\tau=1}^T\sum_{r=1}^{M_\textrm{r}}\textrm{tr}\left(\boldsymbol{\Psi}_r\mathbf{X}[\tau]^H \mathbf{G}_{r}^H\mathbf{G}_r \mathbf{X}[\tau]\right) \\
    &=\sum_{r=1}^{M_\textrm{r}}\sum_{i=1}^{M_\textrm{t}}\sum_{j=1}^{M_\textrm{t}}\Big([\boldsymbol{\Psi}_r]_{i,j}\cdot\sum_{\tau=1}^T\left[\mathbf{X}[\tau]^H \mathbf{G}_{r}^H\mathbf{G}_r \mathbf{X}[\tau]\right]_{j,i}\Big)
    \end{aligned}
    \end{equation}
    Then, for the element at the $i$-th row, $j$-th of $\mathbf{X}[\tau]^H \mathbf{G}_{r}^H\mathbf{G}_r \mathbf{X}[\tau]$, we have
    \begin{equation}\label{OBJ3}
    \begin{aligned}
    &\sum_{\tau=1}^T[\mathbf{X}[\tau]^H \mathbf{G}_r^H\mathbf{G}_r \mathbf{X}[\tau]]_{i,j}\\
    &= \sqrt{\beta_{i,r}\beta_{j,r}}\sum_{\tau=1}^T \mathbf{x}_i[\tau]^H\mathbf{a}_{i,0}\left(\mathbf{U}_i\right)\mathbf{b}^H_{r,0}\mathbf{b}_{r,0}\mathbf{a}^H_{j,0}\left(\mathbf{U}_j\right)\mathbf{x}_j[\tau] \\
    &\overset{(a)}{=} TN\sqrt{\beta_{i,r}\beta_{j,r}}\mathbf{a}^H_{j,0}\left(\mathbf{U}_j\right)\mathbf{E}_j\mathbf{R}\mathbf{E}^H_i\mathbf{a}_{i,0}\left(\mathbf{U}_i\right)
    \end{aligned}
    \end{equation}
    where (a) follows from the definition of $\mathbf{b}_{r,0}$, and the assumption that $\frac{1}{T}\sum_{\tau=1}^T\mathbf{x}[\tau]\mathbf{x}[\tau]^H\approx\mathbf{R}$ when $T$ is sufficiently large \cite{9737357}, $\mathbf{E}_i,1\leq i\leq M_\textrm{t}$ is defined in \eqref{E}. Combining \eqref{OBJ2} and \eqref{OBJ3}, \eqref{OBJ} can be acquired.

   \setcounter{equation}{0}
    \renewcommand\theequation{D.\arabic{equation}}
      \section{Proof of Proposition 4}\label{APP4}
      It can be observed that the objective function $\omega$ \eqref{OBJ} and the power constraint \eqref{C2'} can all be equivalently expressed as a function of the covariance matrix $\mathbf{R}$ of transmitted signal. Therefore, without loss of optimality, we propose to optimize $\mathbf{R}$ instead of $(\mathbf{w},\mathbf{s}_0)$ and formulate the optimization problem of the sensing-only scenario as \eqref{P3} with additional constraint C7 to maintain the positive semi-definiteness of $\mathbf{R}$.
      
      To continue, we consider the case where the optimal solution for $(\mathscr{P}_4)$ is found as $\left(\mathbf{R}', \mathbf{u}'\right)$. Then for arbitrary $\mathbf{u}$ satisfying constraint C3 and C5, it is evident that we can convert $\mathbf{a}_{t,0}\left(\mathbf{U}_t\right)$ into $\mathbf{a}_{t,0}\left(\mathbf{U}'_t\right)$ with a diagonal matrix $\mathbf{D}_t\in\mathbb{C}^{N\times N}$ whose diagonal element $[\mathbf{D}_t]_{n,n}$ satisfying $|[\mathbf{D}_t]_{n,n}|=1,1\leq n\leq N$, i.e., $\mathbf{a}_{t,0}\left(\mathbf{u}'_t\right)=\mathbf{D}_t\mathbf{a}_{t,0}\left(\mathbf{u}_t\right)$. Therefore, with the help of $\mathbf{D}_t,1\leq t\leq{M_\textrm{t}}$, we can have 
      \begin{equation}\label{ADRDA}
      \begin{aligned}    \mathbf{a}^H_{i,0}\left(\mathbf{U}_i\right)\mathbf{D}_i^H\mathbf{E}_i\mathbf{R}'\mathbf{E}_j^H&\mathbf{D}_j\mathbf{a}_{j,0}\left(\mathbf{U}_j\right)=\\
      &\mathbf{a}^H_{i,0}\left(\mathbf{U}'_i\right)\mathbf{E}_i\mathbf{R}'\mathbf{E}_j^H\mathbf{a}_{j,0}\left(\mathbf{U}'_j\right).
      \end{aligned}
      \end{equation}
      According to Appendix \ref{APP3} and \eqref{ADRDA}, given arbitrary transmit FA positions $\mathbf{u}$, the optimal sensing performance can be achieved by $\hat{\mathbf{R}}$ with $\mathbf{E}_i\hat{\mathbf{R}}\mathbf{E}_j=\mathbf{D}_i^H\mathbf{E}_i\mathbf{R}'\mathbf{E}_j\mathbf{D}_j, 1\leq i,j\leq M_\textrm{t}$. It can be easily proved that $\textrm{tr}(\mathbf{E}_t\hat{\mathbf{R}}\mathbf{E}_t^H)=\textrm{tr}(\mathbf{E}_t\mathbf{R}'\mathbf{E}_t^H)$ and $\hat{\mathbf{R}}\succeq\mathbf{0}$. Therefore, constraint $\textrm{C2}'$ and C7 can also be satisfied, which completes the proof.

    \setcounter{equation}{0}
    \renewcommand\theequation{E.\arabic{equation}}
      \section{Proof of Proposition 6}\label{APP6}
      To begin with, $\tilde{\zeta}_k\left(\mathbf{u}_{t,n}\right)$ can be equivalently rewritten as \eqref{til_zeta} shown at the bottom of this page, where $\varrho_{t,n,l_1,l_2}\left(\mathbf{u}_{t,n}\right)\triangleq \vert\left[\mathbf{P}_{t,k,n}\right]_{l_1,l_2}\vert\textrm{cos}\big(\angle\left[\mathbf{P}_{t,k,n}\right]_{l_1,l_2}+\frac{2\pi}{\lambda}(\boldsymbol{\varpi}^\textrm{t}_{t,k,l_1}-\boldsymbol{\varpi}^\textrm{t}_{t,k,l_2})\mathbf{u}_{t,n}\big)$, $\vartheta_{t,n,l}\left(\mathbf{u}_{t,n}\right)\triangleq\vert\left[\mathbf{q}_{t,k,n}\right]_{l,1}\vert\textrm{cos}\big(\angle\left[\mathbf{q}_{t,k,n}\right]_{l,1}+\frac{2\pi}{\lambda}\boldsymbol{\varpi}^\textrm{t}_{t,k,l}\mathbf{u}_{t,n}\big)$.
    \begin{figure*}[b]
	\centering
	\hrulefill
	\vspace*{0pt} 
	\begin{equation}\label{til_zeta}
    \begin{aligned}           \tilde{\zeta}_k\left(\mathbf{u}_{t,n}\right)=&\sum\nolimits_{l=1}^L\left[\mathbf{P}_{t,k,n}\right]_{l,l}+2\sum\nolimits_{l_1=1}^{L-1}\sum\nolimits_{l_2=l_1+1}^{L}\varrho_{t,n,l_1,l_2}\left(\mathbf{u}_{t,n}\right)+2\sum\nolimits_{l=1}^L\vartheta_{t,n,l}\left(\mathbf{u}_{t,n}\right)
    \end{aligned}
	\end{equation}
    \end{figure*}
    Then, $\tilde{\zeta}_k\left(\mathbf{u}_{t,n}\right)$ is reformulated via second-order Taylor expansion as
      \begin{equation}\label{F1}
      \begin{aligned}
         \tilde{\zeta}_k\left(\mathbf{u}_{t,n}\right) &= \tilde{\zeta}_k\left(\mathbf{u}'_{t,n}\right) + \nabla \tilde{\zeta}_k\left(\mathbf{u}'_{t,n}\right)\left(\mathbf{u}_{t,n}-\mathbf{u}'_{t,n}\right)\\
         &+ \frac{1}{2}\left(\mathbf{u}_{t,n}-\mathbf{u}'_{t,n}\right)^T\nabla^2 \tilde{\zeta}_k\left(\boldsymbol{\xi}_{t,n}\right)\left(\mathbf{u}_{t,n}-\mathbf{u}'_{t,n}\right),
      \end{aligned}
      \end{equation}
      where $\boldsymbol{\xi}_{t,n}=\mathbf{u}'_{t,n}+\mu\left(\mathbf{u}_{t,n}-\mathbf{u}'_{t,n}\right)$ with $\mu\in(0,1)$,
      \begin{equation}
      \begin{aligned}
      \nabla \tilde{\zeta}_k\left(\mathbf{u}_{t,n}\right)=\big[\frac{\partial \tilde{\zeta}_k\left(\mathbf{u}_{t,n}\right)}{\partial x_{t,n}},\frac{\partial \tilde{\zeta}_k\left(\mathbf{u}_{t,n}\right)}{\partial y_{t,n}}\big],
      \end{aligned}
      \end{equation}
      where $\frac{\partial \tilde{\zeta}_k\left(\mathbf{u}_{t,n}\right)}{\partial x_{t,n}}$ and $\frac{\partial \tilde{\zeta}_k\left(\mathbf{u}_{t,n}\right)}{\partial y_{t,n}}$ are given by \eqref{par1} and \eqref{par2}, respectively, shown at the bottom of next page, where $\tilde{\varrho}_{t,n,l_1,l_2}\left(\mathbf{u}_{t,n}\right)\triangleq \vert\left[\mathbf{P}_{t,k,n}\right]_{l_1,l_2}\vert\textrm{sin}\big(\angle\left[\mathbf{P}_{t,k,n}\right]_{l_1,l_2}+\frac{2\pi}{\lambda}(\boldsymbol{\varpi}^\textrm{t}_{t,k,l_1}-\boldsymbol{\varpi}^\textrm{t}_{t,k,l_2})\mathbf{u}_{t,n}\big)$, $\tilde{\vartheta}_{t,n,l}\left(\mathbf{u}_{t,n}\right)\triangleq\vert\left[\mathbf{q}_{t,k,n}\right]_{l,1}\vert\textrm{sin}\big(\angle\left[\mathbf{q}_{t,k,n}\right]_{l,1}+\frac{2\pi}{\lambda}\boldsymbol{\varpi}^\textrm{t}_{t,k,l}\mathbf{u}_{t,n}\big)$,
    \begin{figure*}[b]
    \vspace{-12pt}
	\centering
    \hrulefill
	\vspace*{0pt} 
	\begin{equation}\label{par1}
    \begin{aligned}
        \frac{\partial \tilde{\zeta}_k\left(\mathbf{u}_{t,n}\right)}{\partial x_{t,n}}=&-\frac{4\pi}{\lambda}\sum\nolimits_{l_1=1}^{L-1}\sum\nolimits_{l_2=l_1+1}^{L}\tilde{\varrho}_{t,n,l_1,l_2}\left(\mathbf{u}_{t,n}\right)\left[\boldsymbol{\varpi}^\textrm{t}_{t,k,l_1}-\boldsymbol{\varpi}^\textrm{t}_{t,k,l_2}\right]_{1,1} -\frac{4\pi}{\lambda}\sum\nolimits_{l=1}^L\tilde{\vartheta}_{t,n,l}\left(\mathbf{u}_{t,n}\right)[\boldsymbol{\varpi}^\textrm{t}_{t,k,l}]_{1,1}
    \end{aligned}
	\end{equation}

    \vspace{-10pt}
	\begin{equation}\label{par2}
    \begin{aligned}
    \frac{\partial \tilde{\zeta}_k\left(\mathbf{u}_{t,n}\right)}{\partial y_{t,n}}=&-\frac{4\pi}{\lambda}\sum\nolimits_{l_1=1}^{L-1}\sum\nolimits_{l_2=l_1+1}^{L}\tilde{\varrho}_{t,n,l_1,l_2}\left(\mathbf{u}_{t,n}\right)\left[\boldsymbol{\varpi}^\textrm{t}_{t,k,l_1}-\boldsymbol{\varpi}^\textrm{t}_{t,k,l_2}\right]_{1,2}-\frac{4\pi}{\lambda}\sum\nolimits_{l=1}^L\tilde{\vartheta}_{t,n,l}\left(\mathbf{u}_{t,n}\right)[\boldsymbol{\varpi}^\textrm{t}_{t,k,l}]_{1,2}
    \end{aligned}
	\end{equation}
    \end{figure*}
    
     \begin{equation}
     \nabla^2 \tilde{\zeta}_k\left(\mathbf{u}_{t,n}\right)=\begin{bmatrix}\frac{\partial^2 \tilde{\zeta}_k\left(\mathbf{u}_{t,n}\right)}{\partial x_{t,n}^2} & \frac{\partial^2 \tilde{\zeta}_k\left(\mathbf{u}_{t,n}\right)}{\partial x_{t,n}\partial y_{t,n}}\\
     \frac{\partial^2 \tilde{\zeta}_k\left(\mathbf{u}_{t,n}\right)}{\partial y_{t,n}\partial x_{t,n}} & \frac{\partial^2 \tilde{\zeta}_k\left(\mathbf{u}_{t,n}\right)}{\partial y_{t,n}^2}
     \end{bmatrix},
     \end{equation}
      where $\frac{\partial^2 \tilde{\zeta}_k\left(\mathbf{u}_{t,n}\right)}{\partial x_{t,n}^2}$, $\frac{\partial^2 \tilde{\zeta}_k\left(\mathbf{u}_{t,n}\right)}{\partial x_{t,n}\partial y_{t,n}}=\frac{\partial^2 \tilde{\zeta}_k\left(\mathbf{u}_{t,n}\right)}{\partial y_{t,n}\partial x_{t,n}}$ and $\frac{\partial^2 \tilde{\zeta}_k\left(\mathbf{u}_{t,n}\right)}{\partial y_{t,n}^2}$ are given by \eqref{par3}, \eqref{par4} and \eqref{par5}, respectively, shown at the bottom of next page.

    \begin{figure*}[b]
	\centering
    \hrulefill
	\vspace*{0pt} 
	\begin{equation}\label{par3}
    \begin{aligned}
    \frac{\partial^2 \tilde{\zeta}_k\left(\mathbf{u}_{t,n}\right)}{\partial x_{t,n}^2}=&-\frac{8\pi^2}{\lambda^2}\sum\nolimits_{l_1=1}^{L-1}\sum\nolimits_{l_2=l_1+1}^{L}\varrho_{t,n,l_1,l_2}\left(\mathbf{u}_{t,n}\right)\left[\boldsymbol{\varpi}^\textrm{t}_{t,k,l_1}-\boldsymbol{\varpi}^\textrm{t}_{t,k,l_2}\right]^2_{1,1}
    -\frac{8\pi^2}{\lambda^2}\sum\nolimits_{l=1}^L\vartheta_{t,n,l}\left(\mathbf{u}_{t,n}\right)[\boldsymbol{\varpi}^\textrm{t}_{t,k,l}]^2_{1,1}
    \end{aligned}
	\end{equation}

    \vspace{-8pt}
	\begin{equation}\label{par4}
    \begin{aligned}
   \frac{\partial^2 \tilde{\zeta}_k\left(\mathbf{u}_{t,n}\right)}{\partial x_{t,n}\partial y_{t,n}}=-\frac{8\pi^2}{\lambda^2}\sum\nolimits_{l_1=1}^{L-1}\sum\nolimits_{l_2=l_1+1}^{L}\varrho_{t,n,l_1,l_2}\left(\mathbf{u}_{t,n}\right)&\left[\boldsymbol{\varpi}^\textrm{t}_{t,k,l_1}-\boldsymbol{\varpi}^\textrm{t}_{t,k,l_2}\right]_{1,1}\left[\boldsymbol{\varpi}^\textrm{t}_{t,k,l_1}-\boldsymbol{\varpi}^\textrm{t}_{t,k,l_2}\right]_{1,2}\\
   &\qquad\qquad-\frac{8\pi^2}{\lambda^2}\sum\nolimits_{l=1}^L\vartheta_{t,n,l}\left(\mathbf{u}_{t,n}\right)[\boldsymbol{\varpi}^\textrm{t}_{t,k,l}]_{1,1}[\boldsymbol{\varpi}^\textrm{t}_{t,k,l}]_{1,2}
    \end{aligned}
	\end{equation}

    \vspace{-15pt}
	\begin{equation}\label{par5}
    \begin{aligned}
    \frac{\partial^2 \tilde{\zeta}_k\left(\mathbf{u}_{t,n}\right)}{\partial y_{t,n}^2}=&-\frac{8\pi^2}{\lambda^2}\sum\nolimits_{l_1=1}^{L-1}\sum\nolimits_{l_2=l_1+1}^{L}\varrho_{t,n,l_1,l_2}\left(\mathbf{u}_{t,n}\right)\left[\boldsymbol{\varpi}^\textrm{t}_{t,k,l_1}-\boldsymbol{\varpi}^\textrm{t}_{t,k,l_2}\right]_{1,2}^2-\frac{8\pi^2}{\lambda^2}\sum\nolimits_{l=1}^L\vartheta_{t,n,l}\left(\mathbf{u}_{t,n}\right)[\boldsymbol{\varpi}^\textrm{t}_{t,k,l}]^2_{1,2}
    \end{aligned}
	\end{equation}
    \end{figure*}
      
      To acquire a lower bound of $\tilde{\zeta}_k\left(\mathbf{u}_{t,n}\right)$, we let

      \begin{equation}
          \begin{aligned}
              \delta_{t,k,n}&\triangleq\frac{16\pi^2}{\lambda^2}\big(\sum_{l_1=1}^{L-1}\sum_{l_2=l_1+1}^{L}\vert\left[\mathbf{P}_{t,k,n}\right]_{l_1,l_2}\vert+\sum_{l=1}^L\vert\left[\mathbf{q}_{t,k,n}\right]_{l,1}\vert\big).
          \end{aligned}
      \end{equation}
      so that $\delta_{t,k,n}\geq \Vert\nabla^2 \tilde{\zeta}_k\left(\boldsymbol{\xi}_{t,n}\right)\Vert_F$. Since 
      \begin{equation}
      \begin{aligned}
        &\Vert\nabla^2 \tilde{\zeta}_k\left(\boldsymbol{\xi}_{t,n}\right)\Vert_F=\sqrt{\sum\nolimits_i\sigma^2_i\big(\nabla^2 \tilde{\zeta}_k\left(\boldsymbol{\xi}_{t,n}\right)\big)}\\
        &\qquad\qquad\geq\sigma_\textrm{max}\big(\nabla^2 \tilde{\zeta}_k\left(\boldsymbol{\xi}_{t,n}\right)\big)=\Vert\nabla^2 \tilde{\zeta}_k\left(\boldsymbol{\xi}_{t,n}\right)\Vert_2,
      \end{aligned}
      \end{equation}
       where $\sigma_i\big(\nabla^2 \tilde{\zeta}_k\left(\boldsymbol{\xi}_{t,n}\right)\big)$ is the corresponding singular value of $\nabla^2 \tilde{\zeta}_k\left(\boldsymbol{\xi}_{t,n}\right)$, $\sigma_\textrm{max}\big(\nabla^2 \tilde{\zeta}_k\left(\boldsymbol{\xi}_{t,n}\right)\big)$ is the maximal singular value of $\nabla^2 \tilde{\zeta}_k\left(\boldsymbol{\xi}_{t,n}\right)$,we have
       \begin{equation}
      \delta_{t,k,n}\mathbf{I}_2\succeq\Vert\nabla^2 \tilde{\zeta}_k\left(\boldsymbol{\xi}_{t,n}\right)\Vert_2\mathbf{I}_2\succeq\nabla^2 \tilde{\zeta}_k\left(\boldsymbol{\xi}_{t,n}\right),
      \end{equation}
      so that
      \begin{equation}
      \begin{aligned}
          \frac{1}{2}\left(\mathbf{u}_{t,n}-\mathbf{u}'_{t,n}\right)^T\nabla^2 \tilde{\zeta}&\left(\boldsymbol{\xi}_{t,n}\right)\left(\mathbf{u}_{t,n}-\mathbf{u}'_{t,n}\right)\\
          &\enspace\geq -\frac{\delta_{t,k,n}}{2}\left\Vert\mathbf{u}_{t,n}-\mathbf{u}'_{t,n}\right\Vert_2^2.
      \end{aligned}
      \end{equation}
      Above all, the lower bound of $\tilde{\zeta}_k\left(\mathbf{u}_{t,n}\right)$ is thus obtained.

     \end{appendices}

\bibliographystyle{ieeetr}
\bibliography{main}

\vfill

\end{document}